\documentstyle[aps,epsf]{revtex}  % DO NOT DELETE THIS LINE 

\begin{document}        %  DO NOT DELETE OR CHANGE THIS LINE

\baselineskip 14pt
$$\begin{array}{rl}
\hspace{12cm} & \mathrm{CALT\ 68-2220}\\
\ & \mathrm{HUTP-99/A021}\\
\end{array}
$$

\vspace{1cm}
\centerline{\large \bf Charmless Hadronic B Decays at CLEO}

\vspace{0.5cm}

\centerline{Yongsheng Gao}
\centerline{\it Harvard University}

\vspace{0.25cm}

\centerline{Frank W\"urthwein}
\centerline{\it California Institute of Technology}

\vspace{0.5cm}

\centerline{(CLEO Collaboration)}

%\title{Charmless Hadronic B Decays at CLEO}
%\author{Yongsheng Gao}
%\address{Harvard University}
%\author{Frank W\"urthwein}
%\address{California Institute of Technology}
%\author{(CLEO Collaboration)}

% \author{}   % Use this and the next line only if there is a second
% \address{Another University, etc.}  % address. (Remove the left % marks)
%
%\maketitle              % Creates the title area, Do Not Remove

\begin{abstract}        % Do Not Delete this line

The CLEO collaboration has studied two-body charmless hadronic
decays of $B$ mesons into final states
containing two pseudo-scalar mesons, or a pseudo-scalar and a vector meson.
%$h^+h^-$, $h^\pm K^0_S$, and $h^\pm \pi^0$, where $h^\pm$ stands for a charged
%kaon or pion.
We summarize and discuss results presented during the
winter/spring 1999 conference season, and provide a brief outlook
towards future attractions to come.

%present preliminary results based on nearly 6 million $B\bar{B}$~pairs
%collected with the CLEO II and CLEO II.5 detectors.
In particular, CLEO presented preliminary results
on the decays
$B^\pm \rightarrow \pi^\pm\rho^0$
($Br(B^\pm \rightarrow \pi^\pm\rho^0) = (1.5 \pm 0.5 \pm 0.4)\times 10^{-5}$),
$B\rightarrow \pi^\pm\rho^\mp$
($Br(B\rightarrow \pi^\pm\rho^\mp)=(3.5^{+1.1}_{-1.0}\pm 0.5)\times 10^{-5}$),
$B\rightarrow \pi^\pm K^{\star\mp}$
($Br(B\rightarrow \pi^\pm K^{\star\mp}) =
(2.2^{+0.8+0.4}_{-0.6-0.5})\times 10^{-5}$),
and
$B^\pm \rightarrow K^\pm \pi^0$
($Br(B^\pm \rightarrow K^\pm \pi^0) = (1.5 \pm 0.4 \pm 0.3) \times 10^{-5}$)
at DPF99, APS99, APS99, and ICHEP98 respectively.
None of these decays had been observed previously. The first two of these
constitute the first observation of hadronic $b\to u$ transitions.
In addition, CLEO presented preliminary updates on a large number of
previously published branching fractions and upper limits.

%We have also measured
%$Br(B \rightarrow K^\pm \pi^\mp) = (1.4 \pm 0.3 \pm 0.2) \times 10^{-5}$
%and
%$Br(B^\pm \rightarrow K^0\pi^\pm) = (1.4 \pm 0.5 \pm 0.2) \times 10^{-5}$,
%confirming our previously published results. We
%see no significant signal in $B \to \pi^+\pi^-$, and set an upper limit of
%$Br(B \rightarrow \pi^+\pi^-) < 8.4 \times 10^{-6}$ at $90\% $
%confidence level.

\end{abstract}   	% Do Not Delete this line

\section{Introduction and Overview}

  The phenomenon of $CP$ violation, so far observed only in
the neutral kaon system, can be accommodated  by a complex phase in the
Cabibbo-Kobayashi-Maskawa (CKM) quark-mixing matrix~\cite{CKM}.
Whether this phase is the correct, or only, source of $CP$ violation
awaits experimental confirmation.
$B$ meson decays, in particular charmless $B$ meson decays,
will play an important role in verifying this picture.

\begin{figure}[hbp]
\centering
\leavevmode
\epsfxsize=6.5in
\epsffile{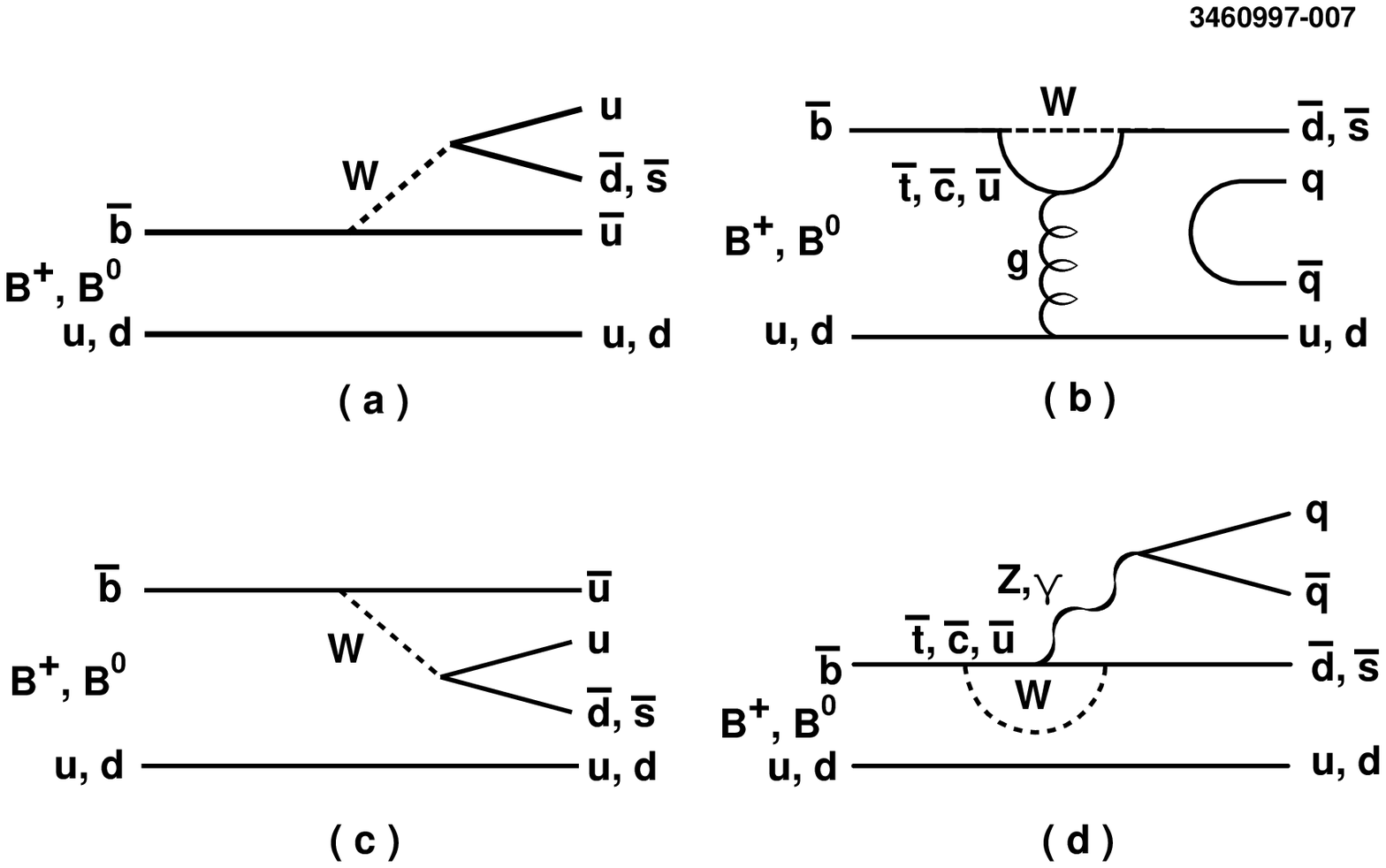}
\caption{The dominant decay processes are expected to be
(a) external W-emission, (b) gluonic penguin, (c) internal W-emission,
(d) external electroweak penguin.}
\label{fig:feynman}
\end{figure}

The decays $B \rightarrow \pi^+\pi^-$ and $B\rightarrow \rho^+\pi^-$,
dominated by the $b \rightarrow
u$ tree diagram (Fig.~\ref{fig:feynman}(a)), can be used to measure $CP$
violation due the interference between $B^0-\bar B^0$\ mixing and
decay. However, theoretical uncertainties due to the presence of
the $b\to dg$ penguin diagram (Fig.~\ref{fig:feynman}(b))
(``{\em Penguin Pollution}'')
make it
difficult to extract the angle $\alpha$ of the unitarity triangle from
$B \rightarrow \pi^+\pi^-$ alone.
Additional measurements of
$B^\pm \rightarrow \pi^\pm \pi^0$, $B,\bar{B} \rightarrow \pi^0\pi^0$,
or a flavor tagged proper-time dependent full Dalitz plot
fit for $B\to\pi^+\pi^-\pi^0$
and the use of isospin
symmetry may resolve these 
uncertainties~\cite{isospin}\cite{grossman_quinn}\cite{snyder-quinn}.
Alternatively, measurement of $CP$ violation due to
the interference of $B\to \pi^\pm\chi_{c0}$ and $B\to\pi^\pm\rho^0$ in
$B^\pm\rightarrow \pi^+\pi^-\pi^\pm$~\cite{bediaga} may provide
information about the angle $\gamma$ of the unitarity triangle.
Neither flavor tagging
nor a measurement of the 
proper-time before decay of the $B$ meson is required
in this case. Extraction of the angle $\gamma$ from this measurement
is subject to theoretical uncertainties due to Penguin Pollution.
However, factorization predicts this to be less severe here than in
$B\to\pi^+\pi^-$ due to at least partial cancelation of the gluonic penguin
contribution among short distance operators with different chirality.

 $B\to K\pi $\ decays are dominated by the
$b \rightarrow sg$ gluonic penguin diagram,
with additional contributions from
$b \rightarrow u$ tree and color-allowed
electroweak penguin (Fig.~\ref{fig:feynman}(d)) processes.
Interference between the penguin (Fig.~\ref{fig:feynman}(b),(d)) 
and spectator (Fig.~\ref{fig:feynman}(a),(c)) amplitudes
can lead to direct $CP$ violation, which
 would manifest itself as a rate asymmetry for
decays of $B$ and $\bar{B}$~mesons.
Several methods of measuring the angle $\gamma$
%,  the phase of $V_{ub}$.
using only decay rates of $B\to K\pi,~\pi\pi$ processes were also
proposed~\cite{triangles}.
This is particularly important, as $\gamma $\ is the
least known parameter of the
unitarity triangle and is likely to remain
the most difficult to determine experimentally.
The ratios
 $R={\cal B}(B\to K^{\pm}\pi^{\mp})/{\cal B}(B^{\pm}\to K^0\pi^{\pm}) $
~\cite{fleischermannel},
and
$R_\star ={\cal B}(B^\pm\to K^{0}\pi^{\pm})/
          2{\cal B}(B^{\pm}\to K^\pm\pi^{0}) $
~\cite{neubert-rosner},
were recently suggested as a way to constrain $\gamma$.
Electroweak penguins and final state interactions (FSI) in
$B\to K\pi$ decays can
significantly affect the former method~\cite{weyers},
whereas the latter method requires knowledge of the ratio $|(T+C)/P|_s$ of
spectator to penguin amplitudes in $b\to s$ transitions.
Uncertainties due to FSI and electroweak penguins are eliminated
using isospin and fierz-equivalence of certain short distance operators.
Studies of $B$ decays to $KK$ final
states can provide useful limits on FSI
effects~\cite{gronaukk}.

$B$ decays to $\eta^\prime K^0_s$, $\rho^0 K^0_s$, and $\phi K^0_s$
may allow for future measurements of $\sin 2\beta$, $\beta$ being the third
angle of the unitarity triangle.
This is of interest because one probes the interference between the amplitudes
for $b\to s$ penguin and $B^0-\bar B^0$\ mixing, rather than
$b\to c$ tree and $B^0-\bar B^0$\ mixing, as done in the more ubiquitous
$B\to\psi K^0_s$ decay. It has been argued~\cite{btosCP}
that a variety of new physics scenarios would affect the CP violating
phase of the $b\to s$ penguin only, leaving the phases of
$B^0-\bar B^0$\ mixing and $b\to u$ tree amplitudes unchanged.
Such new physics scenarios would thus lead to a difference between
proper-time dependent $CP$ violation as measured for example in
$B$ decays to $\eta^\prime K^0_s$ as compared to $B\to\psi K^0_s$.

The present paper presents preliminary CLEO results on
two-body charmless hadronic
decays of $B$ mesons into final states
containing two pseudo-scalar mesons ($B\to PP$),
or a pseudo-scalar and a vector meson ($B\to PV$).
Section II discusses the analysis technique that
is common to all of these analyses.
Results on $B\to PP$ and $B\to PV$ are presented in
Sections III.
Section IV
discusses possible implications of some of the measurements presented.

\section{Data Analysis Technique}

The data set used in this analysis is collected with the CLEO II
and CLEO II.5
detectors at the Cornell Electron Storage Ring (CESR).
%It consists of $5.6~{\rm fb}^{-1}$
Roughly $2/3$ of the data is
taken at the $\Upsilon$(4S)
(on-resonance) while the remaining $1/3$ is taken just
%$2.9~{\rm fb}^{-1}$ taken
below $B\bar{B}$ threshold.
The below-threshold sample
is used for continuum background studies.
The on-resonance sample contains 5.8~million $B\bar{B}$ pairs
for all final states except $\rho^+h^-, K^{\star +}h^-$ ($h^+$ being
a charged kaon or pion), and $\rho^0K^0_s$. For those final states
a total of 7.0~million $B\bar{B}$ pairs was used.
This is an $80\% $ increase in the number of $B\bar{B}$ pairs
over the published analyses~\cite{prl98}.
In addition, we have
re-analyzed the CLEO II data set with improved calibration constants and
track-fitting algorithm allowing us to extend our geometric acceptance
and track quality requirements. This has lead to an overall increase in
reconstruction efficiency of $10 - 20$ $\%$
as compared to the previously published analyses.
The CLEO detector has been decommissioned for a major detector and
accelerator upgrade. Preliminary results based on the full data set of
roughly 10~million $B\bar{B}$ pairs are expected to be ready for the
summer conferences in 1999.

 CLEO II and CLEO II.5 are
general purpose solenoidal magnet detectors, described in
detail elsewhere~\cite{detector}.
 In CLEO II,
 the momenta of charged particles are measured in a
tracking system consisting of a 6-layer straw
tube chamber, a 10-layer precision
drift chamber, and a 51-layer main drift chamber, all operating
inside a 1.5 T superconducting solenoid.  The main drift chamber
also provides a measurement of the specific ionization loss, $dE/dx$,
used for particle identification.
 For CLEO II.5 the 6-layer straw tube chamber was replaced by a 3-layer
double sided silicon vertex detector, and the gas in the main
drift chamber was changed from an argon-ethane to a helium-propane
mixture.
%As a result,
%we have improved our $dE/dx$ and momentum resolution
%for charged kaons
%and pions in $B\to K^\pm\pi^\mp$ by 18\% in CLEO II.5 over CLEO II.
Photons are detected
using a 7800-crystal CsI(Tl) electromagnetic calorimeter. Muons are
identified using proportional counters placed at various depths in the
steel return yoke of the magnet.

Charged tracks are
required to pass track quality cuts based on the average hit residual
and the impact parameters in both the $r-\phi$ and $r-z$ planes.
%Pairs of tracks with vertices
%displaced by at least $3$~mm from the primary
%interaction point are taken as $K_S^0$ candidates.
Candidate $K^0_S$ are selected from pairs of tracks forming well measured
displaced vertices.
%a flight distance significance with respect to the beam spot of at least
%$3\sigma$ for CLEO II and $5.5\sigma$ for CLEO II.5.
Furthermore, we require
the $K^0_S$ momentum vector to point back to the beam spot and
the $\pi^+\pi^-$ invariant mass to be within $10$~MeV, two standard
deviations ($\sigma$),
of the $K^0_S$\ mass.
Isolated showers with energies greater than
$40$~MeV in the central region of the CsI calorimeter
and greater than $50$~MeV elsewhere, are defined to be photons.
Pairs of photons with an invariant mass within $25$~MeV ($\sim 2.5\sigma$)
of the nominal $\pi^0$ mass
are kinematically fitted with the mass constrained to the
$\pi^0$ mass.  To reduce combinatoric backgrounds we
require the
lateral shapes of the showers to be consistent with those from photons.
To suppress further low energy showers from charged particle interactions
in the calorimeter we apply a shower energy dependent isolation cut.

Charged particles are identified as kaons or pions using $dE/dx$.
Electrons are rejected
based on $dE/dx$ and the ratio of the track momentum to the
 associated shower energy in the CsI calorimeter.
We reject muons
by requiring that the tracks do not penetrate the steel absorber to a
depth greater than seven nuclear interaction lengths.
We have studied the $dE/dx$\ separation between
kaons and pions for momenta $p \sim 2.6$~GeV$/c$\ in data using
$D^{*+}$-tagged
$D^0\rightarrow K^- \pi^+$ decays; we find a separation of
$(1.7\pm 0.1)~\sigma$
for CLEO II and $(2.0\pm 0.1)~\sigma$ for CLEO II.5.

We calculate a beam-constrained $B$ mass
$M = \sqrt{E_{\rm b}^2 - p_B^2}$, where $p_B$ is the $B$
candidate momentum and $E_{\rm b}$ is the beam energy.
The resolution in $M$\ ranges from 2.5 to 3.0~${\rm MeV}/{\it c}^2$,
where the
larger resolution corresponds to decay modes with a high momentum $\pi^0$.
We define $\Delta E = \sum_i E_i - E_{\rm b}$, where $E_i$
are the energies of the daughters of the $B$ meson candidate.
The resolution on $\Delta E$ is mode-dependent.
For final states without $\pi^0$'s the $\Delta E$ resolution for CLEO II(II.5)
is $\sim$ 20$-$26(17$-$22)MeV. 
For final states with a high momentum $\pi^0$
the $\Delta E$
resolution is
worse approximately by a factor of two and becomes asymmetric because of
energy loss out of the back of the CsI crystals.
%In $B\rightarrow K^+\pi^0/\pi^+\pi^0$ we parameterize
%the $\Delta E$ resolution
%by a sum of two bifurcated Gaussians of widths $+35/-41(+125/-60)$MeV,
%with 68\% of the total area in the narrower of the two Gaussians.
%The resolution is asymmetric because of energy loss out of the
%back of the CsI crystals.
The energy constraint also helps to distinguish between modes of
the same topology.
For example, $\Delta E$
for  $B \rightarrow K^+ \pi^-$,  calculated assuming
$B \rightarrow \pi^+\pi^-$,
has a distribution that is centered at $-42$~MeV, giving a separation
of $1.6(1.9)\sigma$ between $B \rightarrow K^+ \pi^-$ and
$B \rightarrow \pi^+\pi^-$ for CLEO II(II.5).
In addition, $\Delta E$ is very powerful in distinguishing 
$B\to K^{\star 0}\pi^+$
from $B\to\rho^0\pi^+$, especially if the positive track from the vector
meson is of low momentum.
%Same is true for $B \rightarrow \pi^+ V$ and  $B \rightarrow K^+ V$
%for the $PV$ modes.

We accept events with $M$\ within $5.2-5.3$~$\rm {GeV/c^2}$. 
The fiducial region in $\Delta E$ depends on the final state.
For $B\to PP$ we use $|\Delta E|<200(300)$~MeV
for decay modes without (with) a $\pi^0$\ in the final state. 
The selection criteria for $B\to PV$ are listed in Table II.
This fiducial region
includes the signal region, and a sideband for background determination.

We have studied backgrounds from $b\to c$\ decays and other $b\to u$\ and
$b\to s$\ decays and find that all are negligible for $B$ decays to two
pseudo-scalar mesons.
%The final state mesons have momenta beyond the
%kinematically allowed range for $b\to c$.
In contrast, some of the $B$ decays to a pseudo-scalar and a vector meson
have significant backgrounds from $b\to c$ as well as other charmless
$B$ decays. We discuss these in more detail
below in Section III.
However, the main background in all analyses
arises from $e^+e^-\to q\bar q$\ (where $q=u,d,s,c$).
Such events typically exhibit a two-jet structure and can produce high
momentum back-to-back tracks in the fiducial region.
To reduce contamination from these events, we calculate the angle $\theta_S$
between the sphericity axis\cite{shape} of the candidate tracks and showers
and the
sphericity axis of the rest of the event. The distribution of $\cos\theta_S$\
is strongly peaked at $\pm 1$ for $q\bar q$\ events and is nearly flat
for $B\bar B$\
events. We require $|\cos\theta_S|<0.8$\ 
which eliminates  $83\%$\
of the background
for all final states except those
including $\eta^\prime$ or $\phi$. For the latter final states
a looser cut of $|\cos\theta_S|<0.9$ is used.  

Using a detailed GEANT-based Monte-Carlo simulation~\cite{geant}
we determine overall detection efficiencies (${\cal E}$) ranging from a few
$\%$ to $53\%$ in $B\to K^+\pi^-$.
Efficiencies are listed for all decay modes in the tables in
Section III.
%Efficiencies contain branching fractions for
%$K^0\to K^0_S\to \pi^+\pi^-$\ and $\pi^0\to \gamma\gamma$ where applicable.
We estimate systematic errors on the efficiencies using
independent data samples.

Additional discrimination between signal and $q\bar q$\ background is provided
by a Fisher discriminant technique as described in detail in
Ref.~\cite{phd}.
The Fisher discriminant is a linear combination
${\cal F}\equiv \sum_{i=1}^{N}\alpha_i y_i$\ where the coefficients
$\alpha_i$ are chosen to maximize the separation between the signal
and background Monte-Carlo samples.
The 11 inputs, $y_i$, are $|\cos\theta_{cand}|$ (the cosine of the angle
between the candidate sphericity axis and beam axis), the
ratio of Fox-Wolfram moments $H_2/H_0$~\cite{fox}, and nine variables that
measure the scalar sum of the momenta of tracks and showers from the rest of
the event in
nine angular bins, each of $10^\circ$, centered about the candidate's
sphericity axis. Some of the analyses (final states
including $\eta^\prime$ or $\phi$) use
$|\cos\theta_B|$ (the angle between the $B$ meson momentum and beam axis)
instead of $H_2/H_0$ as one of the inputs to the Fisher discriminant.

We perform unbinned
maximum-likelihood (ML) fits using $\Delta E$, $M$, ${\cal F}$,
$|\cos\theta_B|$ (if not used as input to ${\cal F}$)
and $dE/dx$ (where applicable) as input information for each candidate
event to determine the signal yields.
Resonance masses ($\eta^\prime$ and vector resonances) and
helicity angle of the vector meson are also used 
as input information in the fit where applicable.
%Three different fits are performed, one for each topology
%($h^+h^-$, $h^\pm \pi^0$, and $h^\pm K^0_S $,
%$h^\pm$\ referring to a charged kaon or pion).
In each of these fits the likelihood of the event
is parameterized by the sum of probabilities for
all relevant signal and background hypotheses,
with relative weights determined by maximizing the likelihood function
($\cal L$).
The probability of a particular hypothesis is calculated as a product of the
probability density functions (PDFs) for each of the input variables.
Further details about the likelihood fit can be found in Ref.~\cite{phd}.
%The PDFs of the input variables are parameterized by a Gaussian,
%a bifurcated Gaussian, or
%a sum of two bifurcated Gaussians, except for $|\cos\theta_B|$
%($1-|\cos\theta_B|^2$ for
%signal, constant for background),
%background
%$\Delta E$ (straight line), and background $M$\
%($f(M)\propto M\sqrt{1-x^2}\exp[-\gamma(1-x^2)]$; $x=M/E_b$)
%~\cite{argusbackground}.
%We add a small area ($(3.5\pm 3.5)\% $) Breit-Wigner to
%the background PDF in ${\cal F}$ to conservatively account for
%possible tails in the distribution. This is shown for the $h^+\pi^0$
%final state in Figure~\ref{fig:fdtail}.
The parameters for the PDFs
are determined from independent data and high-statistics Monte-Carlo
samples. We estimate a systematic error on the fitted yield by varying the
PDFs used in the fit within their uncertainties.
These uncertainties are dominated by the limited statistics in the
independent data samples we used to determine the PDFs.
The systematic errors on the measured branching fractions are obtained
by adding this fit systematic in quadrature with the systematic error
on the efficiency.

In decay modes for which we do not see statistically significant yields,
we calculate $90\%$\ confidence level (C.L.) upper limit yields by
integrating the likelihood function
\begin{equation}
{\int_0^{N^{UL}} {\cal L}_{\rm max} (N) dN
\over
\int_0^{\infty} {\cal L}_{\rm max} (N) dN}
= 0.90
\nonumber
\end{equation}
where ${\cal L}_{\rm max}(N)$ is the maximum $\cal L$\ at fixed $N$\ to
conservatively account for possible correlations among
the free parameters in the fit. We then increase upper limit yields by
their systematic errors and reduce
detection efficiencies by their systematic errors to calculate
branching fraction upper limits given in Table I and IV.

\section{Results}

Given the enormous number of results to summarize in this Section, we
choose to show figures only for those decay modes for which 
we observe statistically significant yields, and
no branching fraction measurements have previously been published.
Additional figures for preliminary updates on previously published
branching fraction measurements can be found elsewhere.~\cite{vancouver}

The figures we show are
contour plots of $-2\ln{\cal L}$ for the ML fit
as well as projection plots for some of the fit inputs.
The curves in the contour plots represent
the $n\sigma$ contours , which correspond to the increase
in $-2\ln{\cal L}$ by $n^2$.
Contour plots do not have systematic errors included.
The statistical significance of a given signal yield is determined by
repeating the fit with the signal yield fixed to be zero and recording
the change in $-2\ln{\cal L}$.
For the projection plots we apply additional cuts on all variables
used in the fit except the one displayed.
These additional cuts suppress backgrounds by an order of magnitude at signal
efficiencies of roughly $50\%$.
Overlaid on these plots are the projections of the
PDFs used in the fit, normalized according to the fit results multiplied by
the efficiency of the additional cuts.
All results shown are preliminary. Not all published
analyses~\cite{prl98} have been updated yet.

\subsection{$B$ Decays to Two Pseudo-scalar Mesons}

Table I lists the preliminary 
CLEO results for $B$ decays to two pseudo-scalar mesons. 
Not all possible final states with two pseudo-scalar mesons
have been updated yet. For published results please refer to 
Ref.~\cite{prl98}.

\begin{figure}[htbp]
\centering
\leavevmode
\epsfxsize=3.0in
\epsffile{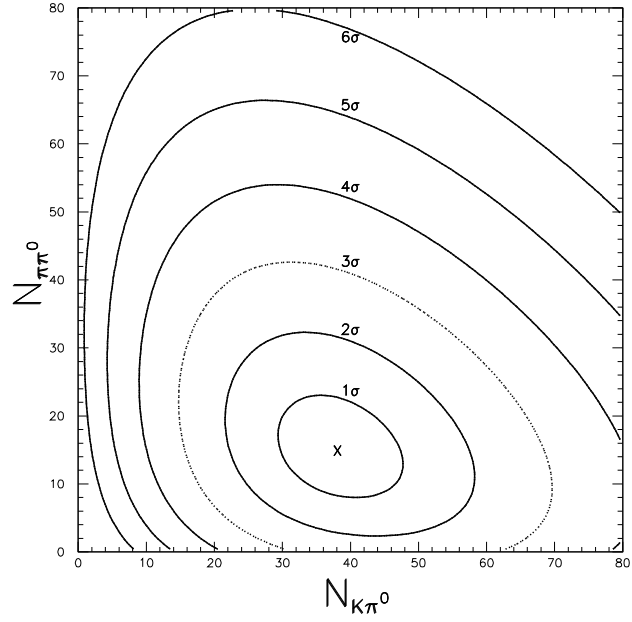}
\caption{Contour of the $-2\ln{\cal L}$ for the ML fit to
$N_{K^\pm \pi^0}$ and $N_{\pi^\pm\pi^0}$ for
$B^\pm \rightarrow K^\pm \pi^0$ and $B^\pm\rightarrow \pi^\pm\pi^0$.
}
\label{fig:contourkpi0}
\end{figure}

\begin{figure}[htbp]
\centering
\leavevmode
\epsfxsize=3.0in
\epsffile{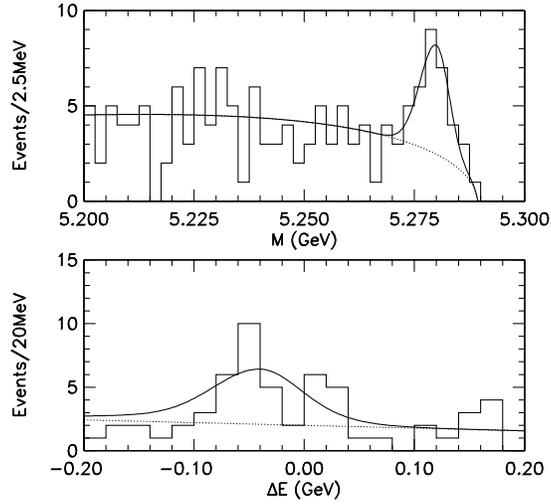}
\caption{Projection plots in $B^\pm \to K^\pm \pi^0$.
$\Delta E$ for $B^\pm\to K^\pm\pi^0$ is centered around $-42$MeV because we
use pion mass when calculating the energy of a track. Only events for
which the candidate track is more likely to be a kaon than a pion 
according to $dE/dx$ enter these figures.}
\label{fig:mass_hpz}
\end{figure}

Figure~\ref{fig:contourkpi0} illustrates a
contour plot for the ML fit to the signal yield ($N$)
in the track $\pi^0$ final state.
The dashed curve marks the $3\sigma$ contour.
To further illustrate the fit,
Figure~\ref{fig:mass_hpz}
shows $M$ ($\Delta E$) projections as defined above.
Events in Figure~\ref{fig:mass_hpz} 
are required to be more likely to be kaons than pions
according to $dE/dx$.
We find statistically significant signals
for the decays $B \to K^\pm \pi^\mp$,
$B^\pm \to K^\pm \pi^0$, $B^\pm \to K^0_S \pi^\pm$, as well as
the two $B\to\eta^\prime K$ decays.
The corresponding branching fractions are listed
in Table~\ref{tab:pp}. Table~\ref{tab:pp} also shows $90\%$ confidence level
upper limits for all the decay modes where we do not measure statistically
significant yields.

\def\red{\ }
\def\black{\ }
\def\blue{\ }

\begin{table}
\begin{center}
\caption{%\Large
Summary of preliminary
CLEO results for $B$ decays to two pseudo-scalar mesons.
Yield and efficiencies in decay modes including $\eta^\prime$
refer to $\eta^\prime\to\eta\pi^+\pi^-, \eta\to\gamma\gamma$
($\eta^\prime\to\rho\gamma$) decays.}
\vskip 0.2cm
\begin{tabular}{lllll}
\hline
Mode & Eff (\%) & Yield & Signif & BR/UL ($10^{-5}$) \\
\hline
%\multicolumn{5}{c}
%{Preliminary Results based on 5.8 million $B\bar B$-pairs:}\\
%\hline
%%%%%%%%%%%
\red{$K^\pm\pi^\mp$}                          & % mode
\red{$53\pm 5$}                                      & % Eff
\red{$43.1_{-8.2}^{+9.0}$}                    & % Yield
\red{$>6\sigma$}                             & % signif
\red{$1.4\pm 0.3 \pm 0.2$}     \\
%%%%%%%%
\red{$K^\pm\pi^0$}                            & % mode
\red{$42\pm 4$}                                      & % eff
\red{$38.1_{-8.7}^{+9.7}$}                     & % Yield
\red{$>6\sigma$}                             & % signif
\red{$1.5\pm 0.4\pm 0.3$}                      \\
%%%%%%%%
\red{$K^0\pi^\pm$}                            & % mode
\red{$15\pm 2$}                                      & % eff
\red{$12.3_{-3.9}^{+4.7}$}                     & % Yield
\red{$>5\sigma$}                             & % signif
\red{$1.4\pm 0.5 \pm 0.2$}    \\
%%%%%%%
\red{$\eta^\prime K^\pm$}                     & % mode
\red{$5(11)$}                                  & % eff
\red{$18.4(50.2)$}                            & % yield
\red{$>6\sigma $}                             & % signif
\red{$7.4^{+0.8}_{-1.3}\pm 1.0$}  \\
%%%%%%%
\red{$\eta^\prime K^0$}                     & % mode
\red{$1.6(3.4)$}                                  & % eff
\red{$5.4(12.7)$}                            & % yield
\red{$>5\sigma $}                             & % signif
\red{$5.9^{+1.8}_{-1.6}\pm 0.9$}  \\
%%%%%%%
\hline
\blue{$\pi^\pm\pi^\mp$}                       & % mode
\blue{$53\pm 5$}                                  & % Eff
\blue{$11.5^{+6.3}_{-5.2}$}                   & % Yield
\blue{$<3\sigma$}                            & % signif
\blue{$<0.84$}\\
%%%%%%%%%%%%%%
\blue{$\pi^\pm\pi^0$}                         & % mode
\blue{$42\pm 4$}                                       & % eff
\blue{$14.9^{+8.1}_{-6.9}$}                   & % Yield
\blue{$<3\sigma$}                            & % signif
\blue{$<1.6$}\\
%%%%%%%
\black{$\eta^\prime \pi^\pm$}                     & % mode
\black{$5(11)$}                                  & % eff
\black{$1.0(0.0)$}                            & % yield
\black{}                             & % signif
\black{$<1.2$}  \\
%%%%%%%%%%%
\hline
%%%%%%%%%%%
\black{$K^\pm K^\mp$}                       & % mode
\black{$53\pm 5$}                                  & % Eff
\black{$0.0_{-0.0}^{+1.6}$}                 & % Yield
                        & % signif
\black{$<0.23$}\\
%%%%%%%%%%%
\black{$K^\pm K^0$}                         & % mode
\black{$15\pm 2$}                                  & % eff
\black{$1.8_{-1.4}^{+2.6}$}                 & % Yield
                         & % signif
\black{$<0.93$}\\
%%%%%%%%
\hline
\end{tabular}
\label{tab:pp}
\end{center}
\end{table}

\subsection{$B$ Decays to a Pseudo-scalar and a Vector Meson}

%Decays of the type $P\to PV$, can be characterized by the helicity angle,
%which is defined (for two-body decays of the vector) to be
%the angle between the direction of a daughter track in the vectors
%rest 
%frame and the direction of the vector in the $B$ candidate's rest frame.
Helicity conservation dictates that the polarization of the vector in
$B\to PV$ is purely longitudinal (helicity = 0 state). 
The kinematics of these decays (assuming two-body decay of the vector)
therefore results in a final state 
with two energetic particles and one soft particle. 
The pseudo-scalar 
$P$ is always very energetic, with a momentum range from 2.3 to 2.8 GeV.
On the other hand
the decay daughters from the vector meson have a very wide momentum range.
While the more energetic particle has momentum between 1.0 and 2.8 GeV,
the soft particle can have momentum as low as 200 MeV.

The backgrounds from $\mbox{$B\bar{B}$}$ events are potentially dangerous
as they may peak in either or both of the $M$ and $\Delta E$ 
distributions.
There are two types of $\mbox{$B\bar{B}$}$ backgrounds that can contribute
to $PV$: $b \rightarrow c$ processes and other rare b processes.

Among the $B \rightarrow P V$ modes we are searching for,
$B\rightarrow \pi^{\pm} V$ and $B\rightarrow K^{\pm} V$ can be well 
separated,
using the dE/dx information of the very energetic $\pi^{+}$ or $K^{+}$
and the separation in $\Delta E$, just like the $B \rightarrow P P$ modes.

Crosstalk of two kinds exist among $PV$ modes. 
First, $\rho\leftrightarrow K^\star$ misidentification is possible for
track $\pi^0$ as well as two track decays of the $\rho$ or $K^\star$
if the fast particle is misidentified due to
the limited particle ID for fast tracks.
Crosstalk among $B^{+}\rightarrow h^{+} \rho^{0}$ and 
$B^{+}\rightarrow h^{+} K^{*0}$ can be controlled to a level
of 20$\%$ or less just by requirements on dE/dx (2 $\sigma$) of the 
decay daughters of the vector meson.
Further separation is achieved by using $\Delta E$ and resonance mass of the
vector as inputs to the likelihood fit.
Second, it is possible to swap a slow momentum pion from the vector
with a slow momentum pion from the other $B$. This is particularly severe
for slow momentum $\pi^0$, as the fake/real $\pi^0$ ratio is about a factor
20 worse for the slow pions than the fast pions from the vector.
In such cases we impose helicity requirement to remove the region with
soft $\pi^0$. Using data (doubly charged vector candidates) and Monte Carlo 
we determine the remaining backgrounds from other rare b processes 
to be small effects that we 
correct for.

The dominant $b \rightarrow c$ background for $PV(h^+ h^+ h^-)$ is 
$B^{+}\rightarrow \bar{D}^0 \pi^{+}$,
where $\bar{D}^0 \rightarrow\pi^{+}\pi^{-}$ or 
$\bar{D}^0\rightarrow K^{+}\pi^{-}$. This particular
background has exactly the same final state particles as 
the $PV(h^+ h^+ h^- )$ signal and therefore peaks
both in $M$ at 5.28 GeV and in $\Delta E$ at 0.0 GeV.
Other $b \rightarrow c$ processes
$B^{+}\rightarrow\bar{D}^0\rho^{+}$ and 
$B^{0}\rightarrow D^{*-}\pi^{+}$ will have peak structure in $M$,
but not in $\Delta E$ due to the missing soft particle.
Because of the large $b \rightarrow c$ branching ratio, the
contribution from these processes needs to be highly suppressed.
We apply a $\bar{D}^0$ (30MeV) veto to all possible $h^{+} h^{-}$ combinations
in $PV(h^+ h^+ h^-)$ modes.

Similarly the $b \rightarrow c$ background for $PV(h^+ h^- \pi^{0}$) are
$B^{0}\rightarrow D^- \pi^{+}$ where $D^{-}\rightarrow \pi^{-}\pi^{0}$
or $B^{0}\rightarrow D^{0} \pi^{0}$ where $D^{0}\rightarrow K^{-}\pi^{+}$.
However, their contribution is negligible due to the branching ratios
involved.

Finally, there are potentially backgrounds from non-resonant
$B$ decays to three-body final states. We test for such backgrounds in data
by allowing a non-resonant signal contribution in the fit, as well as
by determining the fit yield in bins of helicity angle. Neither of these
tests shows any evidence of non-resonant contributions to any of our
final states. The increase of the error on the fitted yield
due to possible non-resonance contributions is accounted for as part of our 
systematic errors.

%The second type of $\mbox{$B\bar{B}$}$ background comes from the
%other rare $b$ ($PV$) processes with different final state particles. 
%For exampke, $PV(h^+ h^- \pi^{0})$ modes for $PV(h^+ h^+ h^-)$ and vice versa.
%When the $\pi^{0}$ is very soft, the two energetic charged particles can 
%pick up another very soft charged particle from the other B and form a fake 
%$PV(h^+ h^+ h^-)$ candidate. These fake $PV(h^+ h^+ h^-)$ events can have 
%peak structure in $M$, while their $\Delta E$ distribution is relatively flat.
%These crossfeed contribution can be easily estimated.

\subsubsection{First Observation of $B^{\pm} \rightarrow \pi^{\pm}\rho^{0}$}

%\subsubsection{Analysis Strategy}

We select separate $h^+ \rho^{0}$ and $h^+ K^{\star 0}$ samples as discussed
above and in Table II.
%by suppressing the 
%contributions from the $B^{\pm}\rightarrow h^{\pm} K^{*0}$ processes as much 
%as possible while keeping efficiencies for 
%$B^{\pm}\rightarrow h^{\pm} \rho^{0}$ modes high.
We then fit for the $B^{+}\rightarrow\pi^{+}\rho^{0}$ and 
$B^{+}\rightarrow K^{+}\rho^{0}$ components in
this $h^{+} \rho^{0}$ sample, as well as a $B\to\pi^+K^{\star 0}$
reflection, averaging over charge conjugate modes.
Similarly we select a $h^{+} K^{*0}$ sample and fit for
the $B^{+}\rightarrow\pi^{+}K^{*0}$ and $B^{+}\rightarrow K^{+}K^{*0}$,
as well as a $B^{+}\rightarrow \pi^{+}\rho^{0}$ reflection.
We do not attempt a simultaneous fit to the
$h^+ \rho^{0}$ and $h^+ K^{\star 0}$ samples
at this point as this would require us to model
the full momentum dependence of $\Delta E$, $dE/dx$, and resonance mass 
in order to separate $\rho$ and $K^*$ contributions.

The variables
$M$, ${\cal F}$,
E($\pi\pi\pi$) $-$ E$_{b}$ (E($\pi K \pi$) $-$ E$_{b}$),
dE/dx of $h$ in $B\rightarrow h \rho^{0}$ ($B\rightarrow h K^{*0}$),
Mass of $\rho^{0}$ ($K^{*0}$) candidate and 
cos($\rho^{0}$ ($K^{*0}$) helicity angle) 
are used to form probability density function (PDF) to perform the ML fit for 
$B^{\pm}\rightarrow h^{\pm} \rho^{0}$
($B^{\pm}\rightarrow h^{\pm} K^{*0}$) sample.
We do not use $dE/dx$ for the daughters of the vector meson in the fit.

%components in this $h^{\pm} K^{*0}$ sample.
%We also include the $B^{\pm}\rightarrow h^{\pm}K^{*0}$ components in the 
%ML fit to the $B^{\pm}\rightarrow h^{\pm} \rho^{0}$ sample 
%(similarly we include the $B^{\pm}\rightarrow h^{\pm} \rho^{0}$ components 
%in the ML fit to the $B^{\pm}\rightarrow h^{\pm} K^{*0}$ sample) to 
%disentangle the fitted yields to get the signal yields for these modes.

\begin{figure}[hbp]
\centering
\leavevmode
\epsfxsize=2.0in
\epsfysize=2.5in
\epsffile{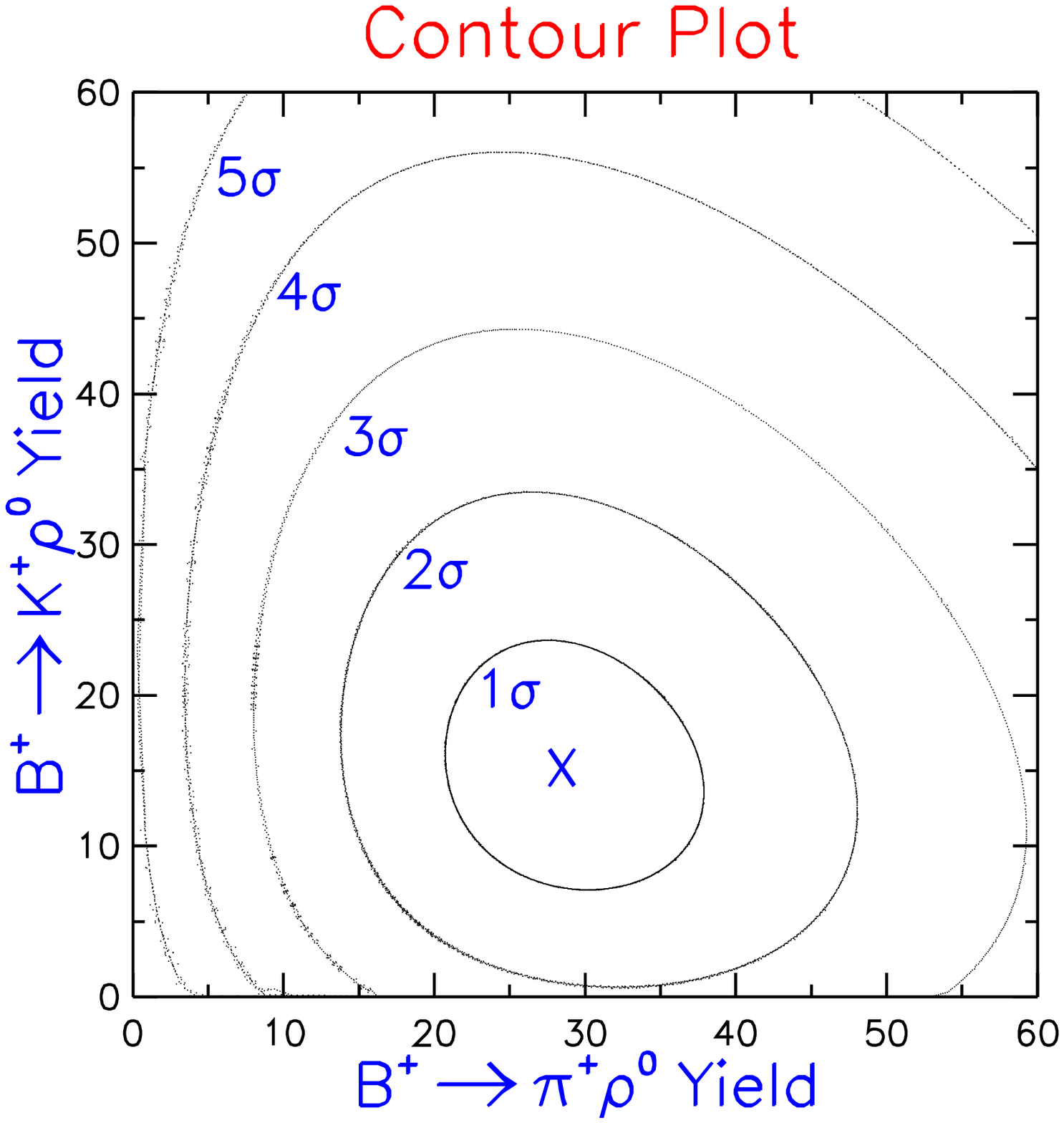}
\epsfxsize=2.0in
\epsfysize=2.5in
\epsffile{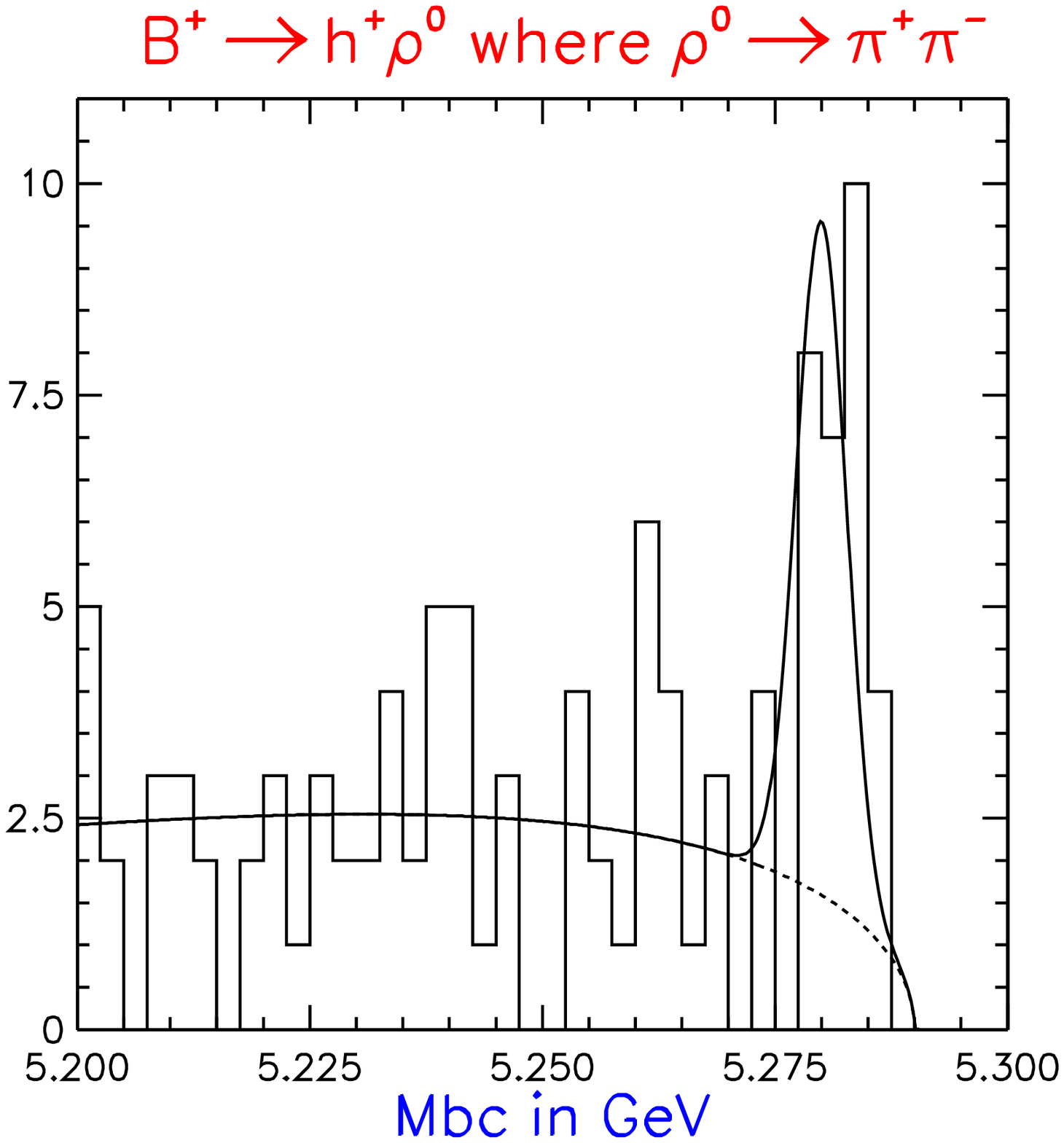}
\epsfxsize=2.0in
\epsfysize=2.5in
\epsffile{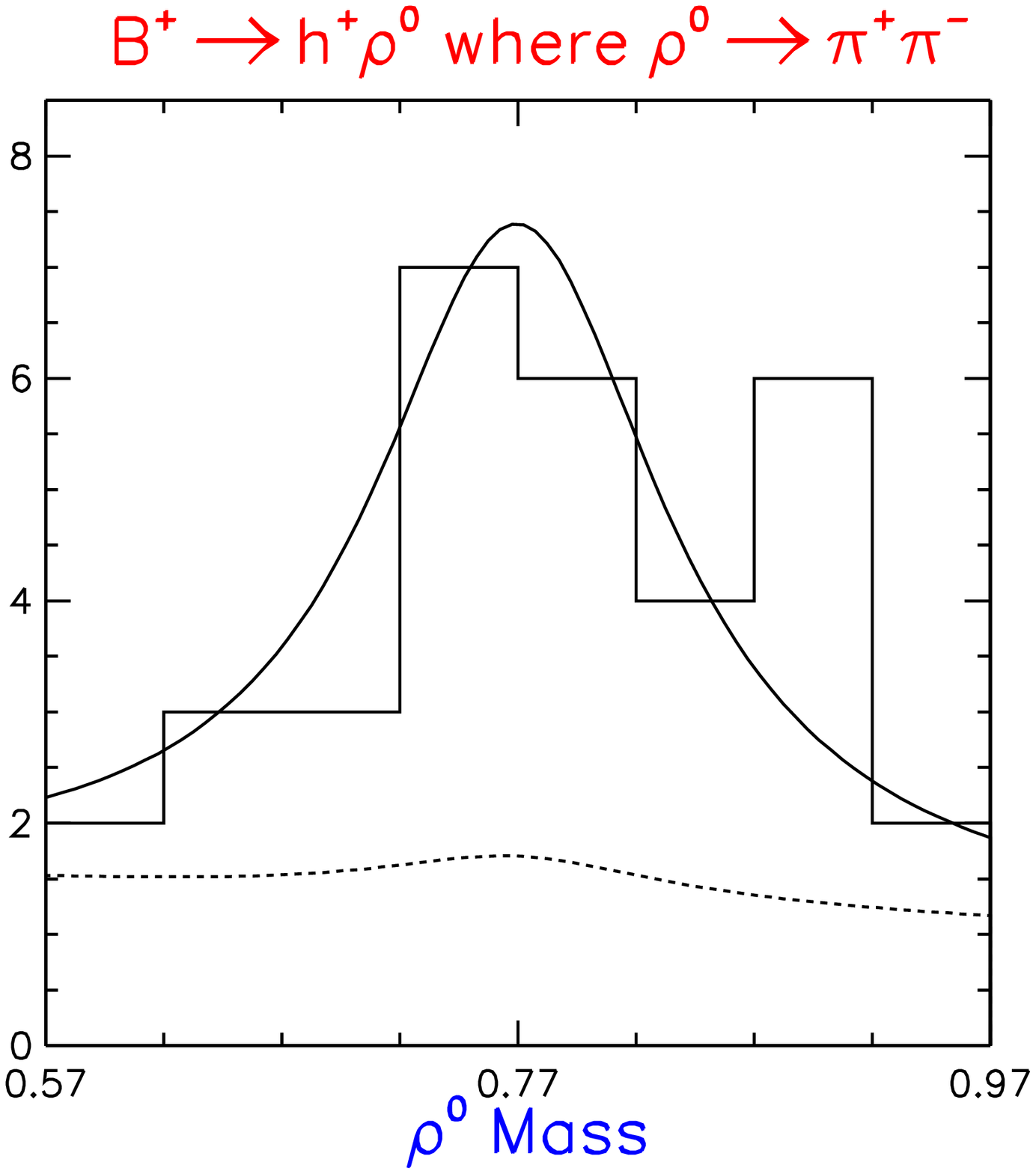}
\caption{Contour and projection plot for 
         $B^{\pm} \rightarrow \pi^{\pm} \rho^{0}$.}
\label{pirho0}
\end{figure}

Efficiencies and results are summarized in Table IV. 
A significant signal in $B^{\pm} \rightarrow \pi^{\pm}\rho^{0}$
is observed. The contour and projection plots are shown in Fig.~\ref{pirho0}.

\vskip 1.0cm

\begin{table}[hhh]
\label{selection}
\caption{Event Selection for $B\to PV$ decays}
 \begin{center}
 \begin{tabular}{|c|c|c|c|} \hline
     Sample           & E $-$ Ebeam         & Resonance Mass Window 
                      & Cos(Resonance Helicity Angle)        
     \\ \hline
 $h^{\pm}\rho^{0}$    & $|$E($\pi\pi\pi$)$-$Ebeam$|<$100MeV       
                      &    200MeV       & $-$0.9 $-$ 0.9       \\ \hline
 $h^{\pm} K^{*0}$     & $|$E($\pi K\pi$)$-$Ebeam$|<$100MeV       
                      &    75MeV        & $-$0.9 $-$ 0.9       \\ \hline
 $h^{\pm}\rho^{\mp}$  & $|$E($\pi\pi\pi^{0}$)$-$Ebeam$|<$150MeV       
                      &    200MeV       &    0.0 $-$ 0.9       \\ \hline
 $h^{\pm} K^{*\mp}(K^{\mp}\pi^{0})$  
                      & $|$E($\pi K\pi^{0}$)$-$Ebeam$|<$300MeV       
                      &    200MeV       &    0.1 $-$ 1.0       \\ \hline
 $h^{\pm} K^{*\mp}(K^0_s\pi^\mp)$  
                      & $|$E($\pi K\pi^{0}$)$-$Ebeam$|<$200MeV       
                      &    200MeV       &    $-$0.86 $-$ 1.0       \\ \hline
\end{tabular}
\end{center}
\end{table}

The contribution of $b\rightarrow c$ and other related rare $B$ processes
are small but not negligible. They are evaluated using about 25 million
generic b$\mbox{$\bar{\mbox{b}}$}$ Monte Carlo events, and specific 
Monte Carlo samples for all the rare 
$B$ processes mentioned in this paper. 
The dominant contributions are listed in Table III. All other contributions
are negligible.
%Central value measurements in \cite{rareb_3} are used to calculate the
%contributions from the related rare $b$ processes.

\begin{table}[hbp]
\label{Correction}
\caption{Contributions to the $B^{\pm}\rightarrow \pi^{\pm}\rho^{0}$ Yield from
                Non-continuum physics backgrounds}
 \begin{center}
 \begin{tabular}{|c|c|} \hline
     Decay Process  
     &   Contribution to $B^\pm\rightarrow \pi^\pm\rho^{0}$ Yield \\ \hline
 b $\rightarrow$ c  &            0.9$\pm$0.7                       \\ \hline
 $B^{0}\rightarrow \pi^{+}\rho^{-}$ 
                    &  0.7$\pm$0.3      \\ \hline
 $B^{0}\rightarrow K^{+}\rho^{-}$  
                    & 0.1$\pm$0.1      \\ \hline
 $B^{0}\rightarrow \pi^{+} K^{*0}$
           & 0.3$\pm$0.2      \\ \hline
% $B^{0}\rightarrow K^{+} K^{*-}$     &   negligible        \\ \hline
 \hline
 TOTAL                &     2.0$\pm$0.8                   \\ \hline
\end{tabular}
\end{center}
\end{table}

The final $B^{\pm}\rightarrow \pi^{\pm}\rho^{0}$ yield after background 
subtraction is: 26.1$^{+9.1}_{-8.0}$ events, leading to a
branching fraction measurement of 
${\cal B}(B^+\to\pi^+\rho^0) = (1.5\pm 0.5\pm 0.4)\times 10^{-5}$.
This is the first observed hadronic $b \rightarrow u$ transition.
%To check whether the signal we observe can come from other rare $b$ processes
%with 3 $\pi$ final state, we exam the Dalitz plot and find no other structures
%besides the $B^{\pm}\rightarrow \pi^{\pm}\rho^{0}$ and 
%$B^{-}\rightarrow D^{0}\pi^{-}$ which does not contaminate into our
%$h^{\pm}\rho^{0}$ sample of ML fit. The helicity and resonance mass 
%distributions are both consistent with coming from 
%$B^{\pm}\rightarrow \pi^{\pm}\rho^{0}$. The result is summarized in
%Table 4. 

\subsubsection{First Observation of $B^{0} \rightarrow \pi^{\pm}\rho^{\mp}$}

As discussed above, the $\pi^0$ daughter of the $\rho$ has a bi-modal
momentum distribution due to the longitudinal polarization 
(helicity = 0) of the $\rho$.
The ratio of real to fake $\pi^0$ is roughly $1/2$ for the low and 
$10/1$ for the high momentum $\pi^0$ region. This leads to largely
increased backgrounds from all sources
as well as multiple entries per event in the
low momentum $\pi^0$ region. In addition, the charged pion tends to be fast
for the slow $\pi^0$ region, thus leading to increased 
$K^{\star +}\leftrightarrow\rho^+$ misidentification.

In contrast, the only drawback of the fast $\pi^0$ region over the three
track sample is a factor two degraded $\Delta E$ resolution.
We therefore choose to use only the half of the sample 
that has a high momentum $\pi^0$ in our fits
in the two track $\pi^0$ final state at this point.
Besides this, the same likelihood fits are made as described for the three 
track final state.

Efficiencies and results are summarized in Table IV. 
The crossfeed rates from other $PV$ modes as well as $b\to c$ decay
backgrounds are negligible. 
A significant signal in $B^{0} \rightarrow \pi^{\pm}\rho^{\mp}$
is observed at a branching fraction of 
${\cal B}(B^0\to \pi^\mp\rho^\pm) = 
(3.5^{+1.1}_{-1.0}\pm 0.5)\times 10^{-5}$.
Note that we do not tag the flavor of the $B$ in the present analysis.
The measured branching fraction is therefore the 
sum of $B^0\to\rho^+\pi^-$ and $B^0\to\rho^-\pi^+$. In addition, averaging 
over charge conjugate states is as always implied.

\vskip 1.0cm

\begin{figure}[hbp]
\centering
\leavevmode
\epsfxsize=2.0in
\epsfysize=2.5in
\epsffile{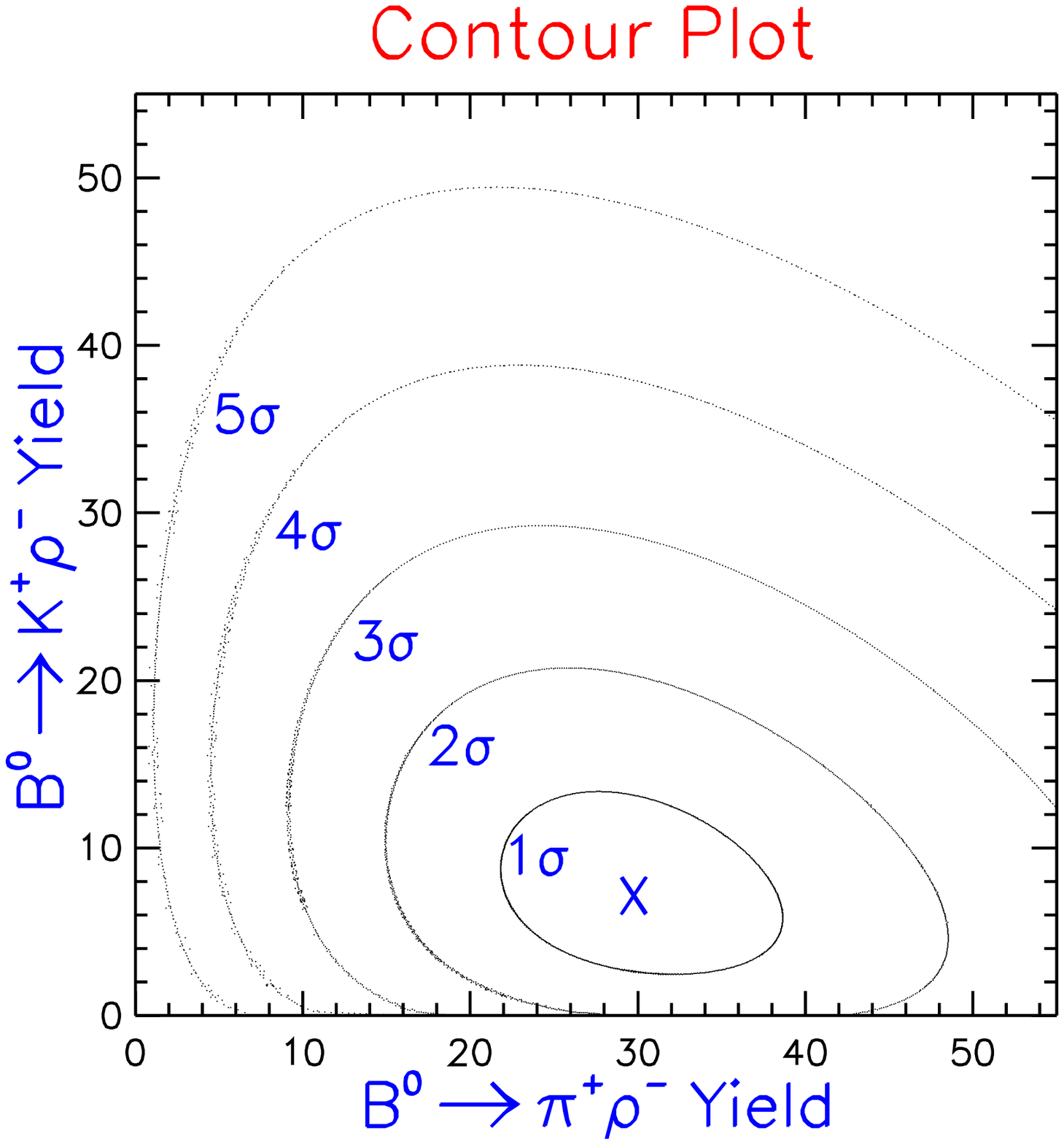}
\epsfxsize=2.0in
\epsfysize=2.5in
\epsffile{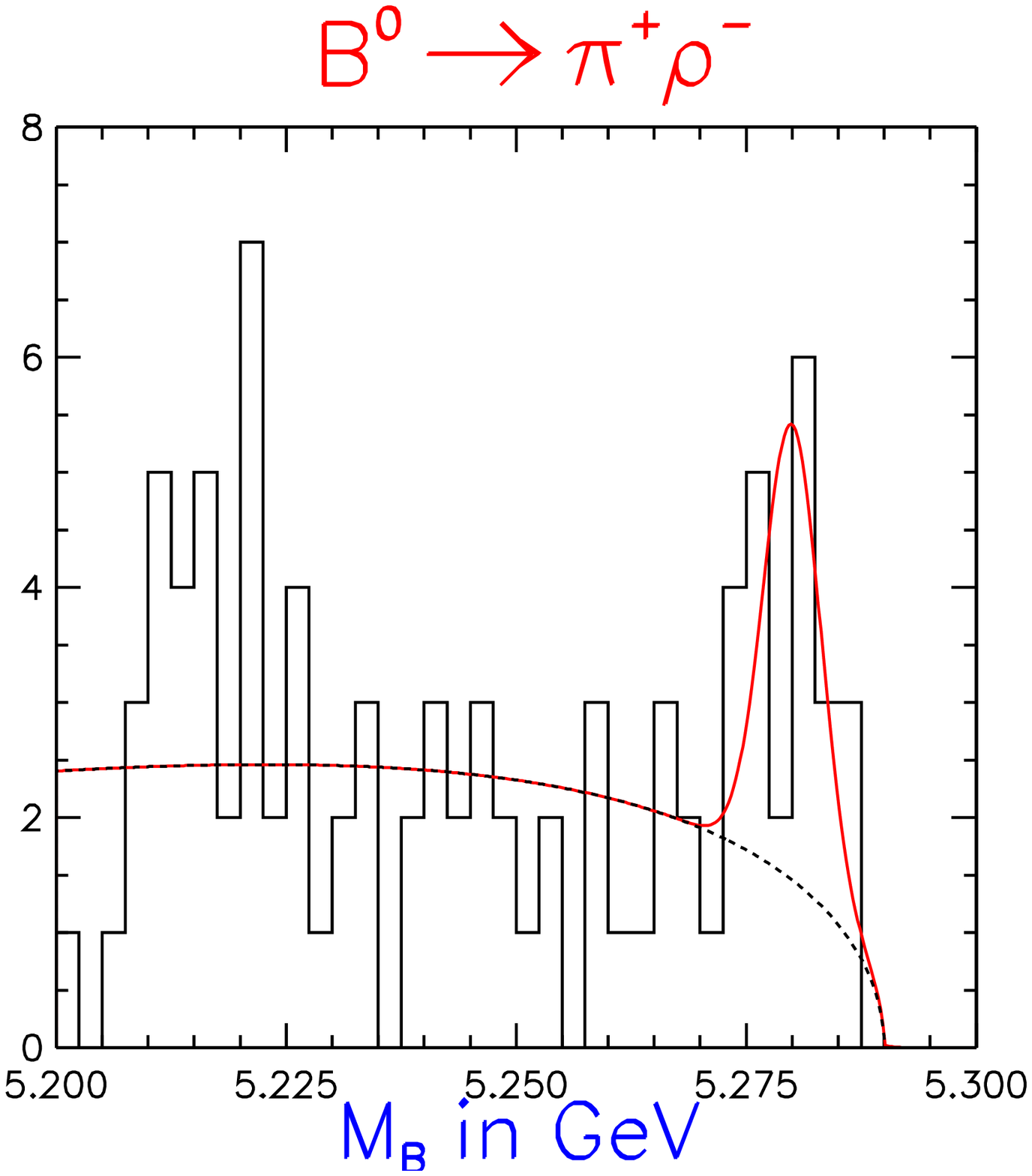}
\epsfxsize=2.0in
\epsfysize=2.5in
\epsffile{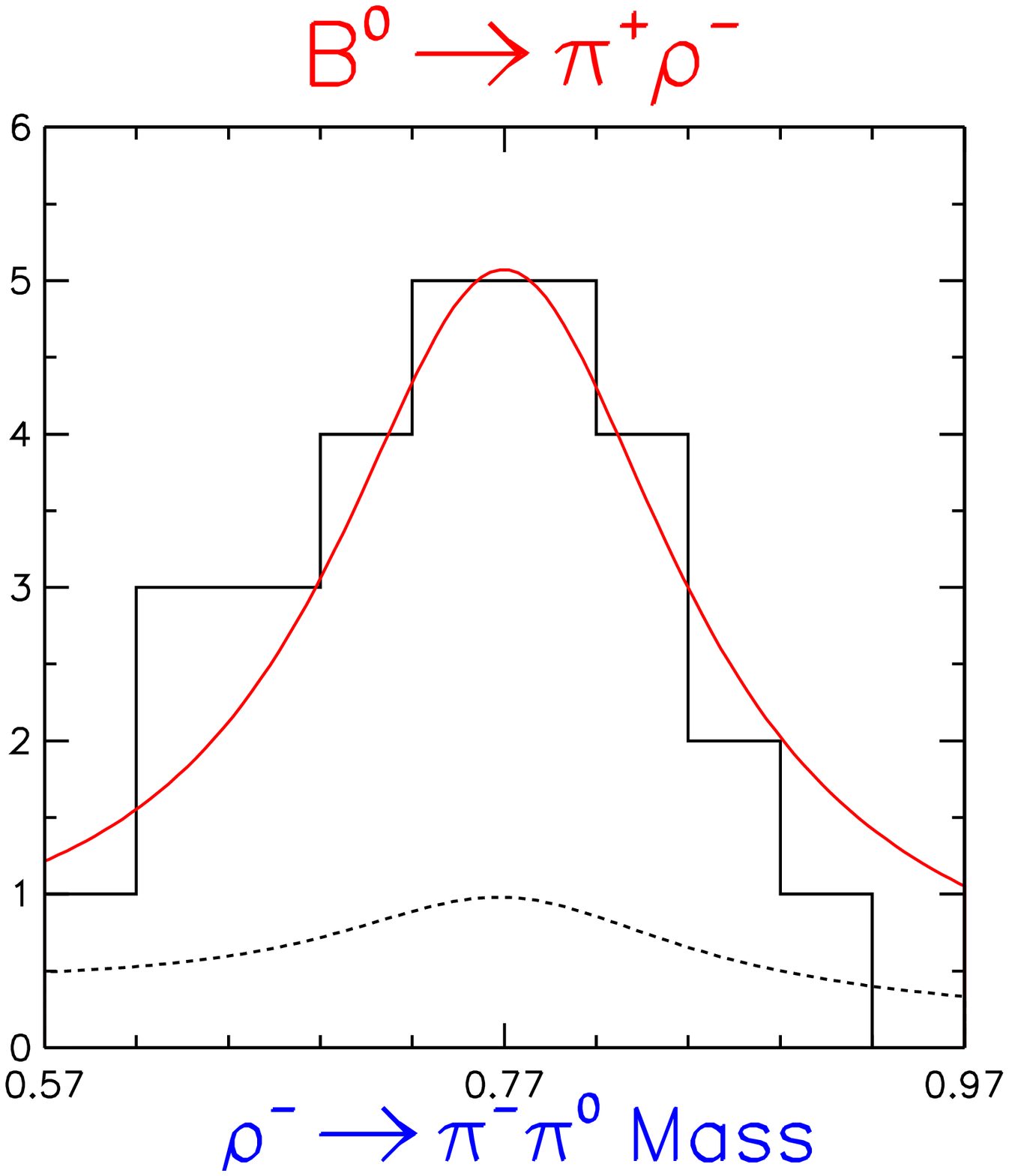}
\caption{Contour and projection plot 
         $B^{0} \rightarrow \pi^{\pm} \rho^{\mp}$.}
\label{pirho}
\end{figure}

%To check whether the signal we observe can come from other rare b processes
%with  $\pi^{+}\pi^{-}\pi^{0}$ final state, we exam the Dalitz plot and 
%find no other structures besides the $B^{0}\rightarrow \pi^{\pm}\rho^{\mp}$ 
%and a few events of possible $B^{0}\rightarrow D^{+}\pi^{-}$ which does not 
%contaminate into the $h^{\pm}\rho^{\mp}$ sample. The helicity and resonance 
%mass distributions are both consistent with coming from 
%$B^{0}\rightarrow \pi^{\pm}\rho^{\mp}$.
%The result is summarized in
%Table X. This is another observed hadronic $b \rightarrow u$ transition.

\subsubsection{Evidence for $B^{0} \rightarrow \pi^{-} K^{*+}$}

We search for $B^{0} \rightarrow \pi^{-} K^{*+}$ with submodes
$K^{*+}\rightarrow K^{0}_{S}\pi^{+}$ and 
$K^{*+}\rightarrow K^{+}\pi^{0}$. 
Due to the large combinatoric and physics backgrounds in the soft $\pi^{0}$
region, we only select the hard $\pi^{0}$ region for the $K^+\pi^0$
decay of the $K^*$. Backgrounds other than those from continuum are
negligible.
Event selections are presented in Table II. Efficiencies and results
are summarized in Table IV. 

\begin{figure}[hbp]
\centering
\leavevmode
\epsfxsize=2.0in
\epsfysize=2.0in
\epsffile{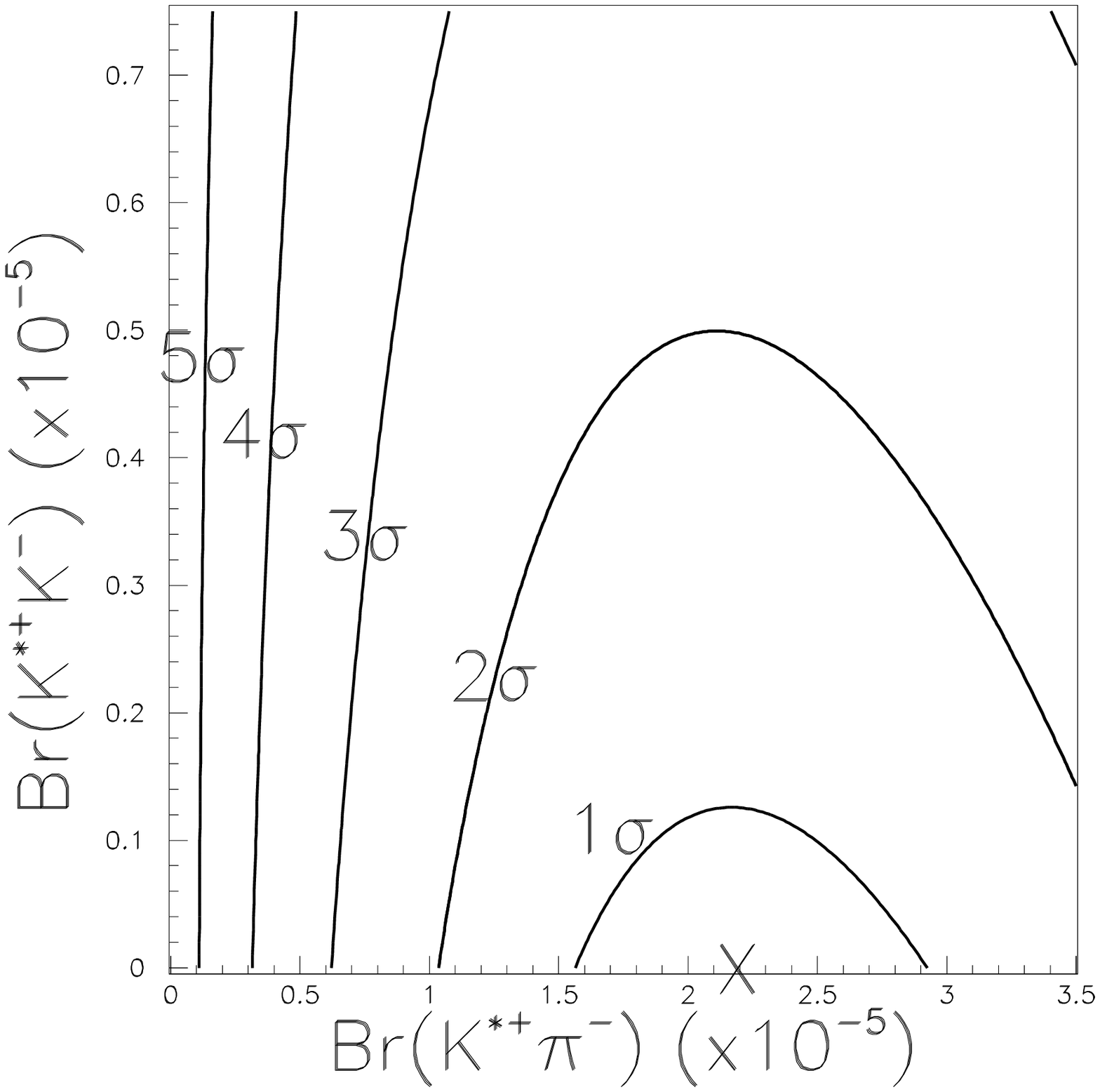}
\epsfxsize=2.0in
\epsfysize=2.0in
\epsffile{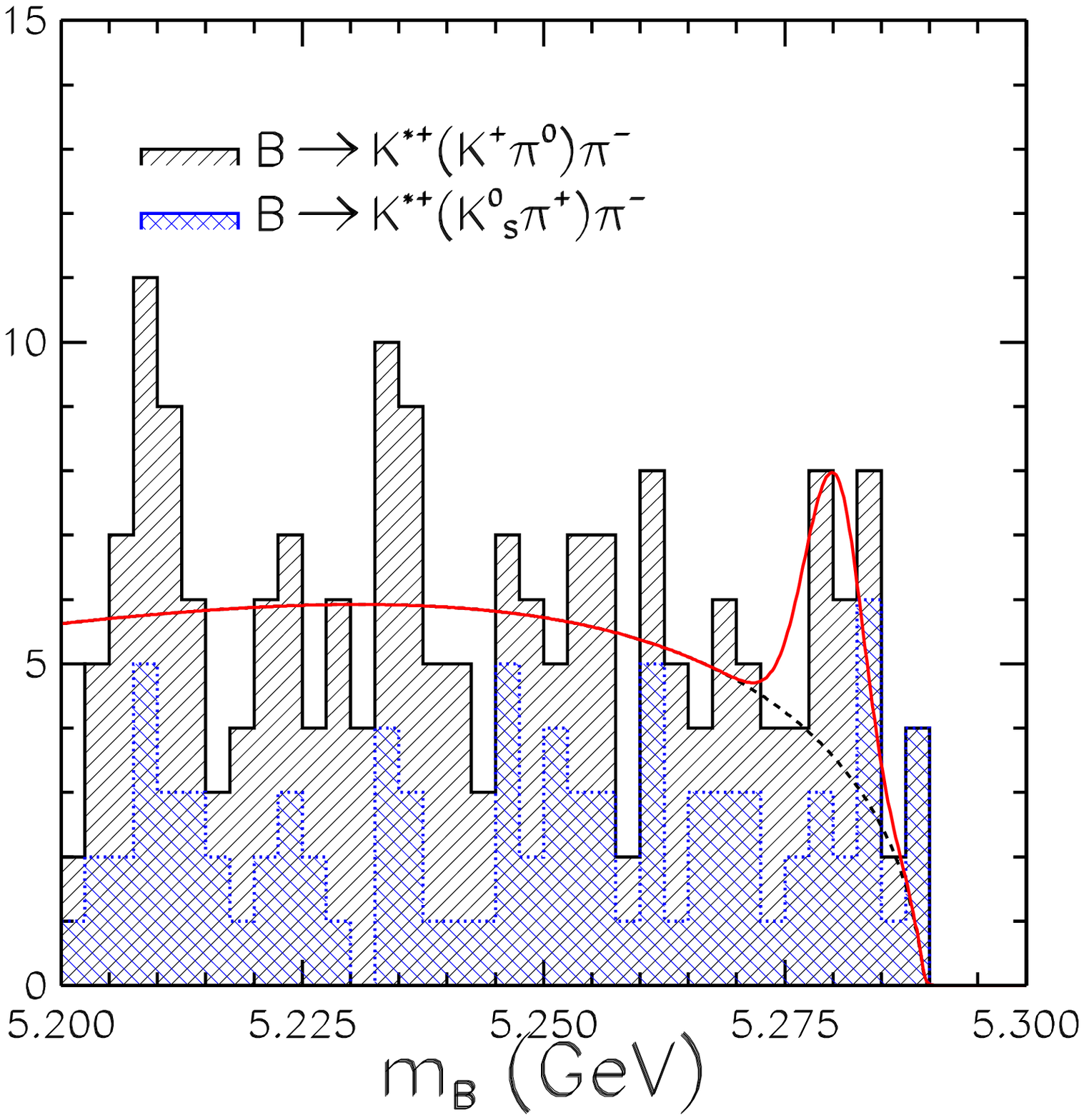}
\epsfxsize=2.0in
\epsfysize=2.0in
\epsffile{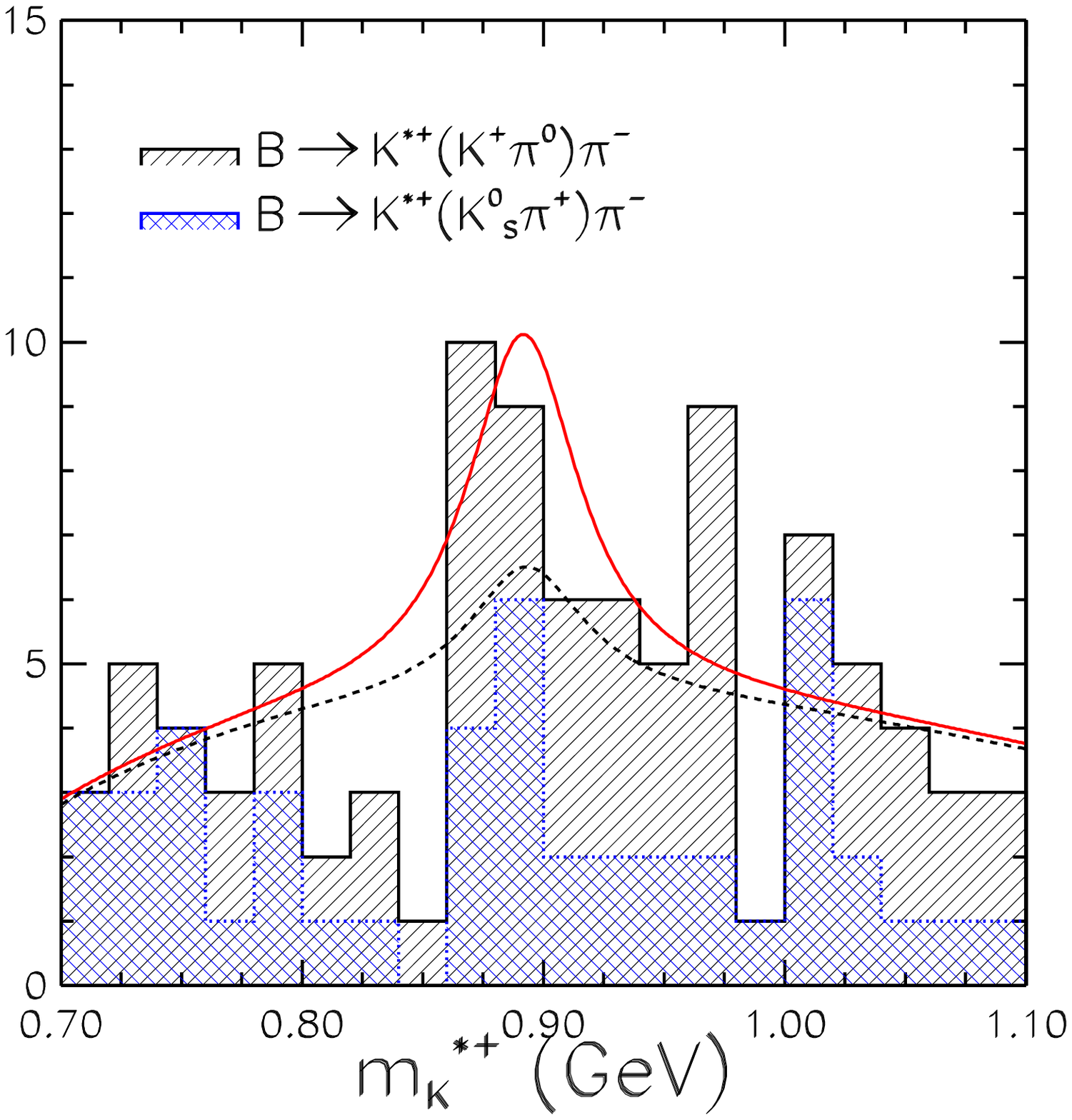}
\caption{Contour and projection plot $B^{0} \rightarrow \pi^{\pm} K^{*\mp}$.}
\label{fig:pikstar}
\end{figure}

The individual branching ratios obtained in the two $K^{*+}$
submodes are consistent, and we combine the two submodes to arrive 
at an average branching ratio of ${\cal B}(B^{0} \rightarrow \pi^{-} K^{*+})$ 
= (2.2 $^{+0.8\ +0.4}_{-0.6\ -0.5}$) $\times$ 10$^{-5}$ which is 5.9$\sigma$
from zero. We note that the statistical significance depends largely
on the two track $K^0_s$ final state, which has less background and
larger efficiency than the two track $\pi^0$ final state.
In contrast to the two observed $B\to\rho\pi$ decays, the one dimensional
projections of the fit (see Fig.~\ref{fig:pikstar})
are somewhat less than inspiring, and a simple event
count in the mass plot does result in an excess of only $2.4\sigma$.
%at a branching fraction one standard deviation lower than the nominal fit.
%of $1.3\pm 0.9\times 10^{-5}$.
%This is explained by three ``golden'' signal events in the two track $K^0_s$
%sample that are somewhat more signal like than 
However, goodness of fit ($21\% CL$), and likelihood per event 
distributions are perfectly consistent with expectations from Monte Carlo.
The most likely signal events have signal 
likelihoods consistent with what one may expect from signal Monte Carlo, 
rather than the background data taken below $B\bar B$ threshold.

In addition,
we generated 25000 distinct Monte Carlo background
samples in the $K^0_s\pi^+\pi^-$ final state. Each of these samples has the
same number of events as our actual data in this final state.
We perform a likelihood fit to each of these 25000 samples 
and record signal yield and significance as reported by each fit.
We find that none of these background samples leads to a reported 
yield or significance as large as found in data. We therefore conclude
that our result is exceedingly unlikely to be due to a background
fluctuation.

\vskip 1.0cm

\begin{table}
\begin{center}
\caption{%\Large
\bf Summary of  CLEO results for $B$ decays to a pseudo-scalar and a 
    vector mesons ($PV$ modes)}
\vskip 0.2cm
\begin{tabular}{lllll}
\hline
Mode & Eff (\%) & Yield & Signif & BR/UL ($10^{-5}$) \\
\hline
%\multicolumn{5}{c}{Preliminary Results based on 5.8 to 7.0 
%million $B\bar B$-pairs:}\\
%\hline
%%%%%%%%%%%
{$\pi^{\pm}\rho^{0}$}                          & % mode
{$30 \pm 3$}                                   & % Efficiency
{$26.1^{+9.1}_{-8.0} $}                        & % Yield
{5.2$\sigma$}                                  & % signif
{$1.5\pm 0.5 \pm 0.4$}     \\
%%%%%%%%
{$\pi^{\pm}\rho^{\mp}$}                        & % mode
{$12 \pm 1$}                                   & % Efficiency
{$28.5^{+8.9}_{-7.9} $}                        & % Yield
{5.6$\sigma$}                                  & % signif
{$3.5^{+1.1}_{-1.0} \pm 0.5$}     \\
%%%%%%%%
{$\pi^{\pm} K^{*\mp}(K^{0}_{S}\pi^{\mp})$}     & % mode
{$7 \pm 1$}                                    & % Efficiency
{$10.8^{+4.3}_{-3.5} $}                        & % Yield
{5.2$\sigma$}                                  & % signif
{$2.3^{+0.9}_{-0.7} \pm 0.3$}        \\
%%%%%%%%
{$\pi^{\pm} K^{*\mp}(K^{\mp}\pi^{0})$}         & % mode
{$4.1 \pm 0.4$}                                & % Efficiency
{$5.7^{+4.3}_{-3.2} $}                         & % Yield
{2.5$\sigma$}                                  & % signif
{$2.0^{+1.5\ +0.3}_{-1.1\ -0.4}$}        \\
%%%%%%%%
{$\pi^{\pm} K^{*\mp}$}                         & % mode
                                               & % Efficiency
                                               & % Yield
{5.9$\sigma$}                                  & % signif
{$2.2^{+0.8\ +0.4 }_{-0.6\ -0.5} $}        \\
%%%%%%%%
{$\pi^\pm K^{\star 0}(K^{+} \pi^{-})$}        & % mode
{$18 \pm 2$}                                   & % Efficiency
{$12.3^{+5.7}_{-4.7} $}                        & % Yield
{$<$ 3$\sigma$}                                & % signif
{$<$ 2.7}     \\
%%%%%%%%
{$K^{\pm}\rho^{0}$}                            & % mode
{$28 \pm 2$}                                   & % Efficiency
{$14.8^{+8.8}_{-7.7} $}                        & % Yield
{$<$ 3$\sigma$}                                & % signif
{$<$ 2.2}     \\
%%%%%%%%
{$K^{0}\rho^{0}$}                              & % mode
{$10 \pm 1$}                                   & % Efficiency
{$8.2^{+4.9}_{-3.9} $}                         & % Yield
{$<$ 3$\sigma$}                                & % signif
{$<$ 2.7}     \\
%%%%%%%%
{$K^{\pm}\rho^{\mp}$}                          & % mode
{$11 \pm 1$}                                   & % Efficiency
{$8.3^{+6.3}_{-5.0} $}                         & % Yield
{$<$ 3$\sigma$}                                & % signif
{$<$ 2.5}     \\
%%%%%%%%
{$K^{\pm}\phi$}                                & % mode
{$26 \pm 3$}                                   & % Efficiency
                                               & % Yield
                                               & % signif
{$<$ 0.59}     \\
%%%%%%%%
{$K^{0}\phi$}                                  & % mode
{$7 \pm 1$}                                    & % Efficiency
                                               & % Yield
                                               & % signif
{$<$ 2.8}     \\
%%%%%%%%
{$\pi^{\pm}\phi$}                              & % mode
{$26 \pm 3$}                                   & % Efficiency
                                               & % Yield
                                               & % signif
{$<$ 0.40}     \\
%%%%%%%%
{$\pi^{0}\phi$}                                & % mode
{$17 \pm 2$}                                   & % Efficiency
                                               & % Yield
                                               & % signif
{$<$ 0.54}     \\
%%%%%%%%
{$K^{\pm} K^{*\mp}(K^{0}_{S} \pi^{\mp})$}       & % mode
{$7 \pm 1$}                                    & % Efficiency
{$0.0^{+0.9}_{-0.0} $}                         & % Yield
                                               & % signif
{$<$ 0.8}     \\
%%%%%%%%
{$K^{\pm} K^{*\mp}(K^\mp \pi^{0})$}       & % mode
{$4.1 \pm 0.4$}                                    & % Efficiency
{$0.0^{+1.3}_{-0.0} $}                         & % Yield
                                               & % signif
{$<$ 1.7}     \\
%%%%%%%%
{$K^{\pm} K^{*\mp}$}       & % mode
                                   & % Efficiency
                         & % Yield
                                               & % signif
{$<$ 0.6}     \\
%%%%%%%%
{$K^{+} K^{*0}(K^{+} \pi^{-})$}                 & % mode
{$18 \pm 2$}                                   & % Efficiency
{$0.0^{+2.1}_{-0.0}$}                                        & % Yield
                                               & % signif
{$<$ 1.2}     \\
\hline
\end{tabular}
\label{tab:pv}
\end{center}
\end{table}

\section{Discussion of our Results}

Let us start by summarizing
some of the more striking features seen in the data.
First of all, we see no evidence for $B\to K\bar K$ decays in either
$B\to PP$ or $B\to PV$.
Such decays would proceed either via highly suppressed $W-$exchange
(e.g. $B\to K^+K^-$) and $b\to d$ penguin diagrams
(e.g. $B\to K^0_sK^\pm$, $B\to K^0_sK^0_s$) or
via final state rescattering (FSI).
Given that our upper limits for some of these decays are an order of
magnitude smaller than at least some of the branching fractions we measure
it seems fair to neglect FSI when trying to understand the
dominant contributions to charmless hadronic $B$ decays.

Second, we see no evidence for $B\to \pi\pi$ decays while we observe both
$B\to K\pi$ as well as $B\to\rho\pi$ decays. We try to make sense out of this
in Section~\ref{ss-model-pp} in the context of isospin and
factorization.

Third, we are so far unable to measure the branching fraction for any of the
$B\to\rho K$ decay modes, despite the fact that we have measured
$B\to\rho\pi$ and $B\to K\pi$, and at least one of the $B\to K^\star\pi$
decay modes. 
This is in full agreement with factorization predictions.
Factorization predicts destructive (constructive) 
interference between penguin operators
of opposite chirality for $B\to\rho K$ ($B\to K\pi$), leading to a rather
small (large) penguin contribution in these decays.
In addition, factorization and CVC predict that only the left-handed
penguin operator contributes in $B\to K^{\star +}\pi^-$. Destructive
interference of penguin operators is therefore not expected in this decay
mode.

Fourth, we want to note that the measured ratio
$R_\rho = {\cal B}(B^0\to\rho^\pm\pi^\mp)/{\cal B}(B^+\to\rho^0\pi^+)$
is much smaller
than naively expected. In $B\to\rho\pi$ decays the $\rho$ can either come
from the upper or lower vertex, and it is generally believed that upper
vertex $\rho$ production clearly dominates due to favorable
form factors as well as decay constants. In addition, $B^+\to\rho^0\pi^+$
is further suppressed by a factor two because only the $u\bar u$ part of the
$\rho^0$ wave function contributes. The present CLEO measurement
of ${\cal B}(B^0\to\rho^\pm\pi^\mp)$ is the sum of upper and lower
vertex $\rho$ production. It is therefore rather surprising that the
measured $R_\rho = 2.3\pm 1.3$ is not significantly larger than two.
Measurements of $B\to\rho^+\pi^0$ as well as a flavor tagged measurement
of $B\to\rho^+\pi^-$ would help to clarify the situation in $B\to\rho\pi$ 
decays. It remains to be seen whether or not such measurements are within reach
using the full CLEO data set.

Finally, maybe the most striking observation in our data are the large
branching fractions measured for charged as well as neutral $B$ decays to
$\eta^\prime K$. 
Violation of a sum-rule proposed by Lipkin~\cite{lipkin} seems to indicate
that a significant flavor singlet contribution is needed to explain
these rates. The literature is full~\cite{etaprime-puzzle} of attempts to
explain this apparent discrepancy, the most interesting of which is the
suggestion that R-parity violating couplings may explain the large 
$\eta^\prime K$ as well as the stringent limit on 
$\phi K$~\cite{rparity}. 
The latter is particularly amusing as one of the relevant couplings
($\lambda^\prime_{323}$) would also be present in 
$B_s-$mixing~\cite{rparity-mixing}
and could therefore lead to a different value for $\gamma$ as inferred from
$B\to K\pi$ decays and the limit on $\Delta m_s/\Delta m_d$ 
in the context of the usual analysis of the $\rho -\eta $ plane
~\cite{rho-eta}.

\subsection{Understanding the non-observation of $B\to\pi\pi$}
\label{ss-model-pp}

Most theoretical predictions lead us to expect a branching fraction for
$B\to\pi^+\pi^-$ at a level of $1-2\times 10^{-5}$.~\cite{all} 
Instead, the central
value and 90$\%$ confidence level upper limit presented here are
$4$ and $8\times 10^{-6}$. With results like this a natural question to ask is
``{\em How small can ${\cal B}(B\to\pi^+\pi^-)$ be?}''.

Let us start our answer by describing a data based factorization prediction.
Assuming factorization, and neglecting W-exchange, penguin annihilation,
and electroweak penguin diagrams one may expect the following
expressions for the decay amplitudes:~\cite{zeppenfeld}

\begin{equation}
\begin{array}{ccccc}
\sqrt{2}A^{\pm 0} &=& -(T+C) \\
A^{+-} &=& -(T+P) &=& -|T|e^{i\gamma}\times (1- |P/T|e^{i\alpha}) \\
\sqrt{2}A^{00} &=& P-C \\
\end{array}
\label{eq:amp}
\end{equation}

Superscripts $+,-,0$ indicate the charge of the final state pions, and
$T,C,P$ stand for external and internal W-emission, and gluonic
penguin diagrams respectively (Fig. 1(a), Fig. 1(c), and Fig. 1(b)).

We can arrive at ``data based factorization estimates'' of these amplitudes
if we identify $C = a_2/a_1 \times T$ and use $a_2/a_1 = 0.21\pm 0.14$
from measurements in $B\to D$
decays~\cite{a2/a1}.
We then estimate $T$
using factorization and the CLEO
measurement
${\cal B}(B\to\pi l\nu) = (1.8\pm 0.5)\times 10^{-4}$~\cite{pilnu}
as follows:

\begin{equation}
\begin{array}{ccccccccc}
T &\sim &
\sqrt{6}\pi f_\pi &
\times &
a_1 &
\times &
\sqrt{d\Gamma(B\to\pi l\nu)/dq^2 |_{q^2 = m_{\pi}^2}
\over \Gamma(B\to\pi l\nu)} &
\times &
\sqrt{{\cal B}(B\to\pi l\nu)} \\
 &\sim & 1.0\mathrm{GeV} &\times &(1.0\pm 0.1) &\times &
(0.27\pm 0.05)/\mathrm{GeV} &\times &
(0.0135\pm 0.0022)\\
 &\sim & \multicolumn{7}{l}{(3.6\pm 0.9)\times 10^{-3}}\\
\end{array}
\label{eq:factorization}
\end{equation}

The dominant error here is due to the spread among a variety of
theoretical models
for the $q^2$ dependence
of the form factor~\cite{lkg}.
We do not assign any error due to a possible breakdown of the
factorization hypothesis.
Throughout this paper we express the absolute size of amplitudes in units of
$\sqrt{\mathrm{Branching\ Fraction}}$.

The decay $B^+\to K^0_s\pi^+$ has three down type quarks in the final state.
Inspection of Figure 1 shows that this final state can only be reached
via penguin diagrams, or final state rescattering. Furthermore, the
electroweak penguin contribution to this decay is color suppressed, rather
than the color allowed one shown in Figure 1(d). It is therefore reasonable
to estimate $P$ from the measured ${\cal B}(B\to K^0\pi^\pm)$
corrected by CKM and SU(3) breaking factors.

Using these numbers we arrive at $|T/P|_d = 5.0\pm 2.3$.
This leads to the factorization predictions
${\cal B}(B^0\to\pi^+\pi^-) = (8\pm 5)\times 10^{-6}$,
${\cal B}(B^+\to\pi^+\pi^0) = (10\pm 5)\times 10^{-6}$,
and ${\cal B}(B^0\to\pi^0\pi^0) \sim $ few $\times 10^{-6}$.
The last of these three estimates is not very meaningful given the errors
on the quantities that enter.
%The error on $B\to\pi^+\pi^-$ includes
%uncertainties due to the relative phase between $T$ and $P$.
%In general, we expect $\sin\alpha >0$, thus leading to destructive
%interference between $T$ and $P$ in $B\to\pi^+\pi^-$. 
We assume maximum
destructive interference
($\cos\alpha = 1$). Ignoring the penguin contribution 
(i.e. $\cos\alpha = 0$) leads
to a prediction of 
${\cal B}(B^0\to\pi^+\pi^-) = (13.0\pm 6.5)\times 10^{-6}$.

As an aside, we can calculate $|T/P|_s = 0.26\pm 0.08$.
This means that CP violating rate asymmetries as large as $50\%$ are
in principle possible for decays like $B\to K^+\pi^-$ if the relevant
weak and strong phases are close to $\pm\pi/2$.

In addition to these factorization estimates, it is quite illustrative
to look at the isospin decomposition of $B\to\pi\pi$:~\cite{isospin}

\begin{equation}
\begin{array}{ccc}
\sqrt{2}A^{\pm 0} &=& 3 A_{3/2}^\gamma\times e^{i\delta}\\
A^{+-} &=& { A_{3/2}^\gamma \times e^{i\delta} + A_{1/2}^\gamma} +
{ A_{1/2}^\beta}  \\
\sqrt{2}A^{00} &=& { 2 A_{3/2}^\gamma \times e^{i\delta} - A_{1/2}^\gamma}
-{ A_{1/2}^\beta} \\
\end{array}
\label{eq:isoamp}
\end{equation}

Here the subscripts $1/2, 3/2$ indicate the
two different isospin amplitudes.
Note that only the $A_{1/2}$ amplitude has
any contribution from $b\to d$ penguins, whereas the $A_{3/2}$ amplitude
is a pure $b\to u$ transition.
We indicate this by making
the dependence on weak ($\beta,\ \gamma$) and
strong interaction phases ($\delta $) explicit.
\footnote{We ignore a possible strong phase difference between penguin and
tree contribution to $A_{1/2}$.}
Using the factorization estimates above, it is easy to show
that $|A_{1/2}^\gamma|,\ |A_{3/2}^\gamma|$, and $|A_{1/2}^\beta|$
are of the same order of magnitude.

Equation~\ref{eq:isoamp} shows that
${\cal B}(B\to\pi^\pm\pi^0)$ can be estimated without making any assumptions
about strong or weak phases.
However very little can be said about the relative size of $B\to\pi^0\pi^0$
versus $B\to\pi^+\pi^-$ without making such assumptions about relative
phases.
Common prejudice assumes
$\delta << 1$ and
therefore $B\to \pi^+\pi^-\ >>\ B\to\pi^0\pi^0$ due to the destructive
interference between $A_{3/2}$ and $A_{1/2}$ in $B\to\pi^0\pi^0$.
However, as we allow for $\delta$ to increase towards $\pi$ we not only
decrease (increase) $B\to\pi^+\pi^- (\pi^0\pi^0)$
but also increase the size of the ``penguin pollution'' in any
future attempt of measuring $\sin 2\alpha$ via time dependent CP violation
in $B\to\pi^+\pi^-$.

We are thus in the amusing situation that we would like $B\to\pi^0\pi^0$
to be large to make the Gronau, London isospin decomposition~\cite{isospin}
experimentally feasible. Though at the same time, we can only hope for
$\delta << 1 $ (i.e. vanishingly small $B\to\pi^0\pi^0$) to avoid destructive
interference between the two $b\to u$ pieces in the amplitude for
$B\to \pi^+\pi^-$.

We conclude that our present data is still
consistent with factorization predictions for $B\to\pi^+\pi^-$.
However, $B\to\pi^+\pi^-$ could be
significantly smaller than predicted by factorization if the strong
interaction phase between isospin amplitudes is non-zero.

\subsection{Comment on Neubert-Rosner bound on $\gamma$}
\label{ss-bound-gamma}

As previously mentioned, the ratio 
$R_\star ={\cal B}(B^\pm\to K^{0}\pi^{\pm})/
          2{\cal B}(B^{\pm}\to K^\pm\pi^{0}) $
~\cite{neubert-rosner},
may be used to constrain $\cos\gamma$ if $R_\star \neq 1$.
Our measurement of this ratio is $R_\star = 0.47\pm 0.24$. 
The relevant equation for bounding $\cos\gamma$ out of the paper
by Neubert and Rosner~\cite{neubert-rosner} is:

\begin{equation}
\cos\gamma = \delta_{EW} - ((1-\sqrt{R_{\star}})/\epsilon_{3/2})/\cos\phi
+ O(\epsilon_{3/2}^2)
\label{eq:eq1}
\end{equation}

The parameter $\epsilon_{3/2}$ is defined in terms of experimentally 
measurable quantities below. It is essentially given by 
the ratio of $b\to u$ tree and
$b\to s$ gluonic penguin amplitudes.
The $O(\epsilon_{3/2}^2)$ terms were shown to be small 
in Ref.~\cite{neubert}. 
Here, $\delta_{EW} = 0.63\pm 0.15$ 
is the theoretically calculated contribution from electroweak penguin
operators~\cite{neubert-rosner}.

The dominant uncertainty in Equation~\ref{eq:eq1} is the unknown
strong phase $\cos\phi$. 
Taking the extreme values of $0$ and $\pi$ for
this phase
we thus arrive at an excluded region for $\cos\gamma$
rather than an actual measurement. To be conservative, one may choose
values for $\delta_{EW}$ such as to minimize this excluded region:

\begin{equation}
0.48 + |1-\sqrt{R_{\star}}|/\epsilon_{3/2} \ge
\cos\gamma \ge  0.78 - |1-\sqrt{R_{\star}}|/\epsilon_{3/2}
\label{eq:bound1}
\end{equation}

The structure of this is obviously to exclude
values for $\cos\gamma$ near
$\cos\gamma = \delta_{EW}$ if 
$X \equiv |1-\sqrt{R_{\star}}|/\epsilon_{3/2}> 0.15$.
The size of the exclusion region is
determined by the central value of $X$
as well as its error.
The variable $X$ defined here 
is given in terms of measurable quantities
up to small uncertainties due to non-factorizable SU(3) breaking:

\begin{equation}\begin{array}{ccc}
X \equiv |1-\sqrt{R_\star}|/\epsilon_{3/2}) & = & |1 - a/b|\times a/c \\
a & =: & \sqrt{{\cal B}(B\to K^0\pi^+)} \\
  & = & (3.74\pm 0.72)\times 10^{-3} \\
b & =: & \sqrt{2\times{\cal B}(B\to K^+\pi^0)} \\
  & = & (5.48\pm 0.91)\times 10^{-3} \\
c & =: & |V_{us}/V_{ud}| f_K/f_\pi \times
         \sqrt{2\times{\cal B}(B\to \pi^+\pi^0)} \\
  & = & (0.95\pm 0.31\ (0.91\pm 0.18))\times 10^{-3} \\
\end{array}
\label{eq:eq2}
\end{equation}

The two different values for $c$ 
are obtained using either 
the most likely value for ${\cal B}(B\to\pi^+\pi^0)$ based on the 
preliminary CLEO results (including statistical and systematic errors added
in quadrature), or a weighted average of the latter with
theoretical predictions based on factorization~\cite{neubert-rosner}.
When calculating $X$ from these numbers
we additionaly increase $c$
to conservatively account for theoretical uncertainties
due to non-factorizable SU(3) breaking
($ ``f_K/f_\pi'' \equiv 1.33$ rather than the experimental value of $1.2$).
The resulting values for $X$ are
$1.15\pm 0.62 $ and $1.20\pm 0.56$ respectively for the two different
values for $c$. In the following we will use $X = 1.20\pm 0.56$.
A number of comments are in order at this point.

First, the central value for $X$ leads to a physical value for
$\cos\gamma$ via Equation~\ref{eq:eq1} only if the strong phase $\phi\sim 0$.
In that case, the measured $X$ then prefers rather large values of
$|\gamma |\sim 120^\circ$.
Such values of $\gamma$ are generally not
the favored ones as they would tend to imply $B_s$ mixing to be smaller
than the present limits and/or 
$f_{B_s}\sqrt{B_{B_s}}/f_{B_d}\sqrt{B_{B_d}}$ to be at the large
end of the generally
assumed range.
It was pointed out by He, Hou, and 
Yang~\cite{hou} that a number of other charmless hadronic $B$ decay
results from CLEO also suggest $|\gamma |> 90^\circ$.

Second, only $10-15\%$ of a Gaussian with mean $0.48 + X$ and 
$\sigma = \sigma_X$ ly within the physically allowed region for $\cos\gamma$.
Calculating a bound in this case isn't all that meaningful. Instead one may
consider $\cos\phi < 0$ to be ruled out at $\sim 90\%$ confidence level.
Using the usual procedure of calculating one-sided 
confidence levels based on the
area inside the physical region only, results in the bound
$\cos\gamma \le 0.33\ @\ 90\%$ confidence level. 

Third, the experimental errors on $X$ are large, 
roughly 1/4 of the physically allowed region total. It is fair to say
that the only reason why we may deduce a non-zero exclusion region
for $\cos\gamma$ from present measurements
is because our present central value for $X$ indicates
a prefered value for $\cos\gamma $ that is far away from 
$\cos\gamma = \delta_{EW}$. 
This is in contrast to some of the recent
analyses of the $\rho - \eta$ plane~\cite{rho-eta} which tend to favor
$\cos\gamma \sim \delta_{EW}$. 

\section{Conclusion}

In summary, we have measured branching fractions for
three of the four exclusive $B\to K\pi$ decays, as well as
the two $B\to\eta^\prime K$ decays,
while only upper limits could be established for all other $B$ decays
to two pseudo-scalar mesons.
In addition, we have observed two of the four $B\to\rho\pi$ decays,
as well as one of the four $B\to K^{\star}\pi$ decays.
We do not observe significant yields for $B$ decays to $\rho K$,
$K^\star K$, $\phi \pi$ or $\phi K$.

The pattern of observed decays is broadly consistent with expectations
from factorization. We see significant contributions from both $b\to u$
as well as $b\to s$ transitions.

In addition,
the Neubert-Rosner bound derived from present CLEO data on charmless
hadronic $B$ decays
indicates $\cos\gamma < 0.33\ @\ 90\%$ confidence level.
This is in slight disagreement with some of the more aggressive analyses of the
$\rho -\eta$ plane found in the literature which prefer larger values of
$\cos\gamma$.

Many thanks to our colleagues at CLEO for many stimulating discussions
as well as the experimental work that made this paper possible.
Further thanks go to A.~Ali, J.-M.~G\'erard, M.~Gronau, W.-S.~Hou, M.~Neubert, 
J.~L.~Rosner, and H.~Yamamoto
for discussions on topics related to Section IV.
We gratefully acknowledge the effort of the CESR staff in providing us with
excellent luminosity and running conditions.
%This work was supported by
%the National Science Foundation,
%the U.S. Department of Energy,
%Research Corporation,
%the Natural Sciences and Engineering Research Council of Canada,
%the A.P. Sloan Foundation,
%the Swiss National Science Foundation,
%and the Alexander von Humboldt Stiftung.

\end{document}